**Comprehensive Lipidomic Automation Workflow using Large Language Models**

Connor Beveridge[1,#], Sanjay Iyer[1,#], Caitlin E. Randolph[1,#], Matthew Muhoberac[1], Palak Manchanda[1], Amy C. Clingenpeel[2], Shane Tichy[3], Gaurav Chopra[1,4]*

[1]Department of Chemistry, 560 Oval Drive, Purdue University, West Lafayette, IN, 47907, [2]ExxonMobil Technology and Engineering Company, Annandale, NJ, 08801; [3]Agilent Technologies Inc. Santa Clara, CA 95051; [4]Department of Computer Science (*by courtesy*), Purdue University, West Lafayette, IN, 47907

[#]These authors contributed equally to this work

*Corresponding Author: gchopra@purdue.edu

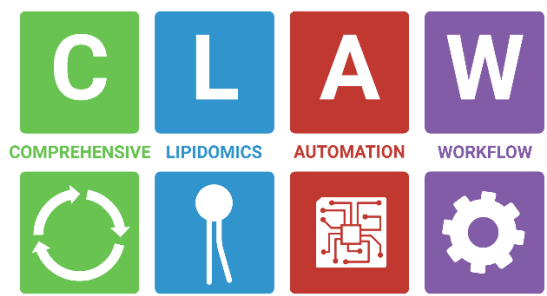

**Abstract**

Profiling lipidome of biological systems generates large amounts of data that makes manual annotation and interpretation time-consuming and challenging. Moreover, the vast chemical and structural diversity of the lipidome compounded by structural isomers further complicates annotation. Although, several commercial and open-source software for targeted lipid identification exists, it lacks automated method generation workflows and integration with existing statistical and bioinformatics tools. We have developed the Comprehensive Lipidomic Automated Workflow (CLAW) platform with integrated workflow for parsing, detailed statistical analysis and lipid annotations based on custom multiple reaction monitoring (MRM) precursor and product ion pair transitions. CLAW is developed with several modules including the ability to identify carbon-carbon double bond position(s) in unsaturated lipids when combined with ozone electrospray ionization (OzESI)-MRM methodology.[1,2] To demonstrate the utility of the automated workflow in CLAW, large-scale lipidomics data was collected with traditional and OzESI-MRM profiling on biological and non-biological samples. Specifically, a total of 1497 transitions organized into 10 MRM-based mass spectrometry methods were used to profile lipid droplets isolated from different brain regions of 18–24 month-old Alzheimer's disease mice and age-matched wild-type controls. Additionally, triacyclglycerols (TGs) profiles with carbon-carbon double bond specificity were generated from canola oil samples using OzESI-MRM profiling. We also developed an integrated language user interface with large language models using artificially intelligent (AI) agents that permits users to interact with the CLAW platform using a chatbot terminal to perform statistical and bioinformatic analyses. We envision CLAW pipeline to be used in high-throughput lipid structural identification tasks aiding users to generate automated lipidomics workflows ranging from data acquisition to AI agent-based bioinformatic analysis.

**Introduction**

Lipids are universal biomolecules that are integral to an array of cellular processes, such as cell signaling, energy conservation, metabolic regulation, and the maintenance of cellular structure.[3,4] In the past decade, there has been a significant increase in lipidomic research due to advancements in mass spectrometry (MS) technologies and recognition within the scientific community that changes in lipid metabolism plays a major role in disease pathologies.[5] For example, chronic diseases, such as Alzheimer's disease (AD), cancer, type 2 diabetes, cardiovascular disease, among others, have shown altered lipid profiles related to disease onset and progression.[6–12] Therefore, detailed lipidome investigations can identify how specific metabolic pathways influences cellular functions and how specific lipids affect health and disease pathogenesis.

Despite the importance of detailed lipidome profiling, the precise characterization, identification, and quantitation of lipids presents significant challenges. MS has emerged as a popular tool in lipidomic analysis, owing to its exceptional sensitivity, specificity, and versatility.[13,14] There exists a set of hierarchical structural information for detailed lipid identification. Lipids can be separated based on class and sum composition that includes the sum of carbon atoms and degree of unsaturation at the lowest hierarchy. The lipid classification system proposed by the LIPID MAPS® consortium classifies lipid structures into eight main categories: fatty acyls (FA), glycerolipids (GL), glycerophospholipids (GP), sphingolipids (SP), sterol lipids (ST), prenol lipids (PR), saccharolipids (SL), and polyketides (PK). Within each lipid category, there exists further structural hierarchical levels including an array of lipid classes and subclasses. Such structural assignments can be readily obtained from conventional MS and tandem-MS (MS/MS) experiments. Independent of ionization mode, lipid precursor ions can be identified at a sum compositional level via accurate mass measurements (i.e., observed mass-to-charge (m/z) ratios). Additional structural information can be obtained by utilizing collision-induced dissociation (CID). For example, individual GP classes like glycerophosphoethanolamines (PEs), glycerophosphoglycerols (PGs), glycerophosphoserines (PSs), glycerophosphoinositols (PIs), glycerophosphocholines (PCs) and sphingomyelin (SM) can be detected and identified based on class-specific fragmentation that relate to GP headgroup composition. Additionally, in negative ion mode, MS/MS of deprotonated acidic GP anions results in the cleavage of ester bonds at the *sn*-1 and *sn*-2 positions of anionic GP. In turn, fatty acyl chains are liberated from the anionic GP precursor ion, yielding abundant carboxylate anions that permit the assignment of fatty acyl sum composition and, in some cases, GP subclass assignment.[15]

Recently, multiple reaction monitoring (MRM)-profiling has demonstrated success for large-scale lipid profiling.[16–18] Briefly, MRM-profiling is a shotgun MS/MS technique for the exploratory analysis of lipids. MRM-profiling facilitates the concurrent analysis of up to thousands of individual precursor/product ion pair transitions in a single sample injection, allowing for the efficient generation of large amounts of lipidomic data that should be analyzed using statistically robust methods to identify up- and down-regulated lipids.[17–19] MRM transitions can be predicted and subsequently generated by exploiting lipid class-specific fragmentation patterns in conjunction with lipid database information. In general, the precursor ion is the expected ionized *m/z* of the lipid molecule at its class or species level, while the product ion is the expected *m/z* for a headgroup or fatty acyl neutral loss from the lipid precursor ion.[20] The combination of the precursor and product ion pair constitutes the MRM transition. Although MRM-profiling is limited to exploratory analysis, the identified lipid targets and are in good agreement with LC-MS/MS lipidomics profiling results.[21,22]

While conventional tandem-MS experiments including MRM-profiling are effective, they have inherent limitations specifically related to isomeric lipids resolution varying in carbon-carbon double bond position and geometry. We note that while basic lipid identification using lipid class and sum composition identifiers can provide substantial biological insights,[17,18] deep lipid structural identification is required to fully understand lipid roles, behavior, identify new targets and biomarkers in metabolic and disease state perturbations. For example, lipid identification at the carbon-carbon double bond (C=C) level has proven useful in differentiating between cancerous and healthy cells.[23,24] In addition, it was demonstrated that only quantitative analysis of lipid C=C location was sufficient to discriminate gefitinib-resistant cells from a population of gefitinib-sensitive cells. [16,21,25–27]

To aid in isomeric resolution, MS-based methods are often coupled with chromatography, novel-ion activation techniques, and chemical conjugation strategies.[28–39] Ozone electrospray ionization (OzESI) and ozone induced dissociation (OzID) have proven highly successful for the detailed identification of unsaturated lipid structures.[2,31,40–42] Briefly, OzESI-MS permits the elucidation of C=Cs within unsaturated lipid structures using ozone-induced fragmentation with the source of a conventional ESI mass spectrometer. In-source ozone can be generated either via (1) a corona discharge at the ESI capillary in an oxygen atmosphere or via (2) an external ozone generator. Upon exposure to ozone, unsaturated lipid ions will undergo facile cleavage at each C=C position. Thus, OzESI-MS results in chemically induced fragment ions that reveal C=C position(s). While effective and easy to implement on commercial platforms, OzESI-MS relies on ozone exposure prior to mass selection. In turn, lipidomics datasets can quickly become difficult to interpret. Therefore, interfacing with liquid chromatography (LC) can reduce complexity in these datasets, but manual interpretation can still be overwhelming and laborious, especially when developing high throughput workflows. In turn, large-scale lipidomics dataset processing, such as those generated by MRM-profiling, are inherently prone to human error and bias.

Currently, comprehensive solutions and standards for MRM-lipidomics data analysis have yet to be established. Various public tools exist that cater to specific parts of the lipidomics data collection and analysis workflow, but overall, the process remains largely manual.[43–49] Moreover, there is no standardized data processing and statistical analysis workflow for large-scale lipidomics data analysis which encompasses parsing raw data files, peak annotation, lipid identification, robust statistical and pathway analysis. Consequently, variations in data processing pipelines and choice of statistical analysis leads to vastly different interpretations and presentations of lipidomics data. In addition, one of the largest challenges faced by modern lipidomics is to keep informatic pipelines sustainable, adaptable, and reproducible, given the rapidly changing landscape of MS-based lipidomics technologies. Therefore, it is essential that future lipidomics analysis pipelines are reliable and robust, and incorporate updated lipidomics standardization, such as suggested nomenclature and data presentation guidelines.

In this work, we introduce an end-to-end integrated and automated MRM lipidomics platform called Comprehensive Lipidomic Automation Workflow (CLAW). The CLAW platform encompasses automated worklist generation for data acquisition (**Figure 1a**), raw data parsing and annotation (**Figure 1b**) to statistically correct differential lipid analysis and identification of related genes in lipid pathways (**Figure 1c**). We aim to bridge current gaps in MS-based lipidomics to develop automated and modular solutions ranging from data collection to processing that are easily extendable to different MS-based methodologies and instruments. Our platform automates worklist generation, lipid annotation with structural details (i.e., C=C localization) and statistically robust bioinformatic analysis. In addition, we introduced a Language User Interface (LUI) and custom AI agents to assist users in proper data collection, annotation, and interpretation based on large language models.[13,14] To demonstrate the versatility of CLAW platform, we report results by using traditional MRM-profiling and online LC-OzESI-MRM profiling to differentiate lipid profiles from different sample types, including lipid droplet extracts from brains of mice and canola oil samples.

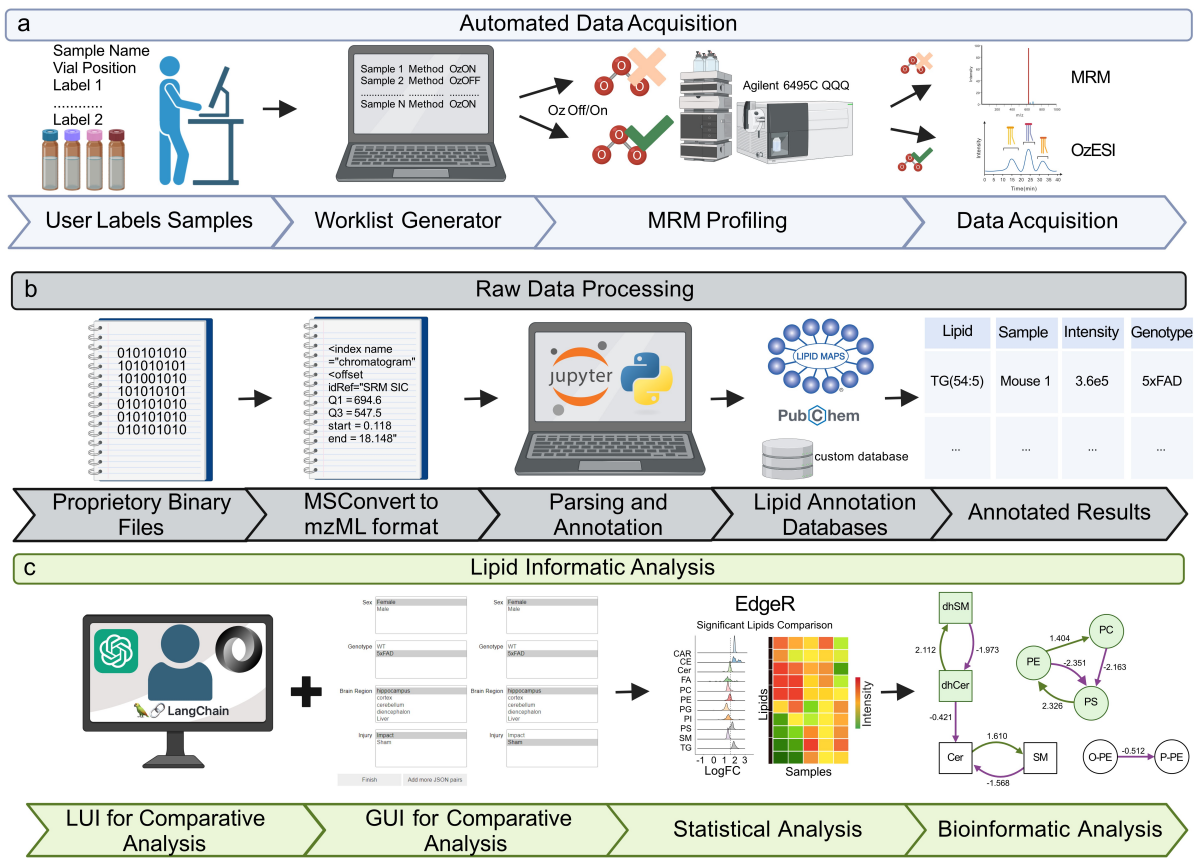

**Figure 1. CLAW MRM lipidomics platform overview.** a) *Automated Data Acquisition:* Using an Agilent 6495C QQQ instrument labeled samples are formatted by the worklist generator and pasted into the MassHunter worklist. After the MRM experiment, proprietary binary (.d) files are exported. b) *Data Processing*: The proprietary binary files are converted to mzML using MSConvert. A script utilizing the pymzml package reads the mzML files, which are then parsed and annotated with a custom lipid database, and then stored in a pandas dataframe. c) *Lipid Informatic Analysis*: A Language User Interface (LUI) and Graphical User Interface (GUI) are developed to provide the user with the ability to compare data from different samples. These interfaces help the user perform robust statistical analysis to visualize and identify significantly expressed lipids. Finally, tools for identifying lipid-based pathways are provided for target gene identification for biological samples.

## Methods

*Nomenclature*

We adopt the nomenclature and shorthand notation as proposed by Liebisch et al. for LIPID MAPS®.[50] Briefly, GP classes include phosphatidylcholine (PC), phosphatidylethanolamine (PE), phosphatidylglycerol (PG), phosphatidylinositol (PI), and phosphatidylserine (PS). Fatty acyl substituents are described with the total number of carbon atoms and double bonds before and after the colon, respectively. If known, identified double bond position(s) are indicated with parentheses following the shorthand notation for FA sum composition. For example, 18:1(9) represents an 18-carbon chain with 1 double bond between carbon-9 and carbon-10 as numbered from the carboxylate moiety.[44,50] In some cases, isomeric FAs are referred to using the "*n*-x" nomenclature where the unsaturation site occurs "x" carbons away from the terminal, i.e. methyl end of the aliphatic chain. To describe triacylglycerol (TG) structure, we first describe TG sum composition, reflecting the sum of carbon atoms and double bonds within the three FA chain substituents. For example, TG 54:2 indicates a TG containing 3 acyl chains whose sum composition adds up to 54 total carbons and 2 carbon-carbon double bonds. To further describe TG structures obtained using traditional-MRM profiling, the following nomenclature is used to express TG sum compositions carrying specific FA chain substituents. For example, TG 54:2_FA 18:1 indicates a TG molecular species with sum composition of 54:2 that carries at least one 18:1 acyl chain. In the case of OzESI-MRM profiling experiments, confirmed double bond location in a given FA substituent is indicated as TG 54:2_FA 18:1*n*-9.

## Mass Spectrometry

### Traditional MRM experiments

MRM experiments were conducted on an Agilent 6495C triple quadrupole mass spectrometer equipped with a jet stream technology ion source (AJS) that has been modified to perform OzESI.[51] Samples were introduced via flow injection (i.e., no chromatography) using an Agilent 1290 II Infinity LC system. For traditional MRM-profiling experiments, dried lipid extracts were dissolved in 200 uL of 1:1 methanol:chloroform with 10 mM ammonium formate. Prior to analysis, reconstituted lipid extracts were diluted 2-fold in 70:30 methanol:acetonitrile with 10 mM ammonium formate. Injection solvent without lipids was used as the 'blank' sample, while injection solvent containing EquiSPLASH™ LIPIDOMIX® (Avanti® Polar Lipids, Alabaster, AL, USA) at a concentration of 0.02 µg/mL was used as a quality control sample to monitor instrument status throughout the run. Briefly, 8 µL of diluted lipid extract was delivered to the AJS source of the mass spectrometer using an Agilent 67167B autosampler. MRM methods were established for 11 lipid classes, covering approximately 1500 individual species tabulated in **Table S1**.

### LC-OzESI-MRM experiments

An online LC-OzESI-MRM method was established to enable unsaturated lipid identification at the C=C level.[51] To facilitate OzESI, high concentration ozone generated via an external generator was delivered to the nebulizer on the Agilent 6495C mass spectrometer. Several canola oil samples were examined for triacylglycerol (TG) content using the developed OzESI-MRM pipeline. Crude, degummed and refined, bleached and degummed (RBD) canola oil samples were diluted 10,000-fold in 1:1 methanol: isopropyl alcohol (IPA). TGs were first separated chromatographically. Briefly, 4 µL of the diluted oil samples was injected onto an EclipsePlus C18 RRHD (1.8 µm, 2.1 x 100 mm) column held at 50°C. To separate TG species based on equivalent carbon number (ECN), we used a gradient consisting of A = IPA with 10 mM ammonium formate and B = acetonitrile (ACN) with 10 mM ammonium formate with a flow rate of 0.6 mL/min. The gradient progresses linearly from 20% A to 60% A over the course of 24 min. The mobile phase composition is held at 60% A for 6 min, before returning to initial conditions (20% A) for another 5 min before terminating the run.

The mass spectrometer was operated in MRM mode. Custom MRMs and OzESI-MRMs were established for unsaturated TG species. Briefly, traditional MRMs monitored the neutral loss of designated acyl chains from ammoniated TG adduct ions. For example, to screen for TG species containing an 18:1 fatty acyl chain, traditional MRMs monitor the neutral loss of 299.2 Da from the [TG + NH$_4$]$^+$ precursor ion.

OzESI-MRMs were established using previously tabulated neutral losses that are indicative of C=C position formed via the reaction of the unsaturated lipid ion with ozone. In general, in-source ozonolysis results in the efficient formation of product ions termed OzESI-aldehydes, reflecting decomposition of the unsaturated lipid at the C=C location and resulting in formation of the diagnostic aldehyde product ion. Ultimately, the investigation of "ozone on" and "ozone off" mass spectra were analyzed by the CLAW automated informatic pipeline for detailed lipid molecular identification, including acyl chain composition and the localization of unsaturation sites. In the "ozone on" experiments, we note that ozonolysis is occurring prior to mass selection. To ensure accurate lipid identification, retention times were matched to correlate precursor and product ion pairs between the "ozone on" and "ozone off" experiments.

*Brain Region-specific Lipid Droplet Sample Preparation*

Wild-type mice (C57BL/6J) and Alzheimer's disease mice (5xFAD) were obtained from the Jackson Laboratory and maintained in a pathogen-free facility. All experiments, including breeding, involving mice were performed in accordance with Purdue University's Institutional Animal Care and Use Committee (IACUC) guidelines. Specifically, 18-24-month-old 5xFAD (n=5 mice) and wild-type mice (n = 3 mice) were perfused, and the brain tissues were isolated and snap-frozen in dry ice chilled iso-pentane and stored at -80 °C. Prior to processing, the brain tissue was divided into four regions (hippocampus, cortex, cerebellum, and diencephalon). Next, the lipid droplets (LDs) from different brain regions were isolated using a lipid droplet isolation kit from Cell Biolabs (San Diego, CA) according to the manufacturer's instructions and then stored at −80 °C. All LD samples were processed together for lipid extraction using the Bligh and Dyer protocol[3]. Briefly, the frozen LDs were thawed in 4°C and 450 µL of cold methanol and 250 µL of chloroform was added. The samples were mixed and vortexed for 10 seconds, resulting in a one-phase solution, which was incubated at 4°C for 15 mins. Next, 250 µL ultrapure water and 250 µL chloroform was added, followed by centrifugation at 16,000 x *g* for 10 mins, resulting in three phases in the tubes. The bottom organic phase that contains the lipids was transferred to new tubes and evaporated in a speed-vac, leaving behind the dried lipid mixture.

**CLAW Description and Operation**

*Python Environment Setup*

The YAML file located in the 'Requirements' folder at https://github.com/chopralab/CLAW.git details the Python environment packages required for CLAW. To install these packages, navigate to this folder and execute in the terminal `pip install -r CLAW.yml`. Once installed, activate this environment in the Jupyter notebook prior to executing. The CLAW Workflow Example is detailed in **Supporting Information**.

*Automatic Worklist Generation*

CLAW is designed to automate data acquisition (**Figure 1a**), data processing (**Figure 1b**) and statistical and bioinformatic analysis (**Figure 1c**). For more efficient worklist creation, CLAW incorporates an automated worklist generator for MRM lipidomics experiments. By automating the worklist construction process, CLAW eliminates the slow and error-prone process of manual worklist creation. For completeness, a detailed tutorial of worklist setup and application example is provided in **Supporting Information**. A Jupyter notebook processes input from three comma separated values (CSV) files and outputs a worklist as a csv file. The first CSV file (Path_info.csv) contains file path data, while the second CSV file (methods.csv) contains a list of MS methods to be used in the worklist. The third CSV file (labels.csv) contains sample name and position (**Figures S1-S6**). Additional columns can also be added to describe sample-specific information which are used by the GUI or LUI for filtering in comparative analysis. After uploading the designated CSV files, the generator then produces an acquisition worklist in the form of a CSV file. The generated worklist CSV file can then be copied and pasted into Agilent MassHunter acquisition software worklist tab (build 10.0.142). Next, the user can initiate data acquisition for the MRM experiments. Although the program has been tested on the Agilent MassHunter software, the worklist generator can be used with other MS acquisition software that support copy and paste functions. We believe that CLAW's automatic worklist generator will reduce preparation time and prevent user errors.

*Data Parsing and Annotation*

Currently, a significant portion of data analysis for different lipidomics methods are processed manually, which is tedious, time-consuming, and error prone. This becomes especially relevant for large data sets that are often obtained from MRM-profiling experiments in a high-throughput manner. To reduce analysis time and help provide more consistent, accurate results, CLAW automates the majority of the MRM-based lipidomics data processing pipeline. Following data acquisition, raw data files are exported and prepared for the CLAW data processing step: parsing and annotation. To prepare data files for parsing in CLAW, raw data files converted to mzML format using MSConvert.[52] Converting the proprietary binary (.d) file format into a more universally accessible mzML format facilitates several downstream applications.

The entire parsing and annotation script is implemented in a Jupyter notebook. Data parsing uses the pymzml (version 2.5.2) package to extract MRM transitions and MS ion count based on the user selected MRM transition database. The lipids are identified based on given precursor and product ion pair transitions as defined by a user-constructed MRM database. The selected database, uploaded as a CSV or excel file, must contain known lipids, classes, and corresponding transitions. An example database with the number of MRM transitions for each lipid class with respective abbreviation is shown in **Table S1**. Potential LIPID MAPS® ID's (LM_IDs) are also assigned if naming convention is compatible with LIPID MAPS® or PubChem through their respective RESTful APIs.

The parsing script operates by matching the precursor and product ion data from the MRM database with the information present in the mzML files. However, it is possible that the *m/z* for each MRM transition value may not be an exact match when rounding to 1 decimal place. To make CLAW capture slight *m/z* differences across low to high-resolution instrumentation, a resolution tolerance parameter is selected that takes the absolute difference between two values allowing non-identical transitions to be matched. Since mass spectrometry has a high selectivity, the default tolerance is set to *m/z* 0.1 but may be adjusted per user specifications.

Each matching transition is annotated with the corresponding lipid from the MRM database and stored in a pandas (version 1.5.2) dataframe. Pandas dataframes offer flexible and intuitive data structures with comprehensive built-in functions, making data manipulation and analysis streamlined. Their compatibility with various data formats, readable syntax, and integration with other key Python libraries further enhance their ease of use.

*Automated OzESI-MRM Prediction and Annotation*

CLAW's parsing and annotation is designed for both direct infusion and LC raw data files. To identify lipid structures with C=C specificity, an LC-OzESI-MRM approach was developed in-house.[51] Briefly, in the LC-OzESI-MRM method, data processing follows the same parsing and annotation procedure as the standard MRM method, with addition of several key steps.

To establish lipid retention times, "ozone off" experiments were first employed. Using a Jupyter notebook interface, a detailed table is compiled to capture all identified lipids along with their corresponding retention times. These retention times serve as ground truth for "ozone on" experiments. Next, OzESI-MRM transitions for identified lipids are automatically generated and tabulated for each LC-OzESI-MS/MS experiment with user-defined sample names and groups. Next, the data from LC-OzESI-MRM "ozone on" experiments are parsed and matched with previously tabulated OzESI-MRM predictions to annotate C=C position for each lipid molecule based on user-specified tolerance in a pandas dataframe with the associated lipid identity and C=C position.

A Gaussian mixture model is used to cluster the data and identify distinct peaks and shapes for each group. The clustering is centered around a retention time window, which is user-defined and based on the ground truth retention times derived from the respective "ozone off" experiments. The cluster that most closely aligns with this ground truth retention time is retained, while other clusters corresponding to the same group are excluded from the annotated pandas dataframe. For each lipid, defined by a unique MRM transition and retention time, the maximum MRM intensity value from the retained cluster is selected and stored. Additionally, users can also visualize chromatograms for specific lipids to validate peak identification and clustering with an option to adjust the retention time window, as needed. Next, the maximum intensity value for each lipid is used to compute relative quantification of unsaturated C=C isomeric ratios based on the double-bond locations specified by the user. These ratios are also determined manually using peak areas from Agilent MassHunter software for comparison.

*Graphical User Interface*

A graphical user interface (GUI) is developed to interactively select lipid MRM data for comparison. The GUI is built using Jupyter notebook (IPython) widgets and is freely available in CLAW's GitHub repository. The GUI filters are applied to the pandas dataframe prior to performing statistical analysis. The filters are stored as JavaScript Object Notation (JSON) files. In general, JSON files are an organized, hierarchical file format that permit filter information saved as key-value pairs to be easily accessed. One of the useful features of CLAW's GUI is the ability to create multiple filters sequentially for datasets with several parameters simultaneously. Specifically, **Figure S7** shows an example where a user can simultaneously filter data based on genotype, sex, brain region, or any combination, to create two JSON files allowing for multiple statistical comparisons to identify differential lipids. The GUI displays parameters from the labels.csv file, previously created at the worklist generation step. The user can continuously generate comparisons by selecting the 'Add More JSON Pairs' button and choosing new parameters. Selections are finalized by clicking the 'Finish' button. Finally, the generated JSON files are used in the Jupyter notebook for data analysis.

*Language User Interface (LUI)*

Another option in CLAW for interacting with data is a language user interface (LUI), which interfaces large language models (LLMs) with custom action agents to interact with various components of the workflow. The LUI uses a chatbot style interface which can provide text-based assistance to a user. Specifically, the LUI interacts with tailor-made AI agents to perform functions. Each AI agent uses the LLM to dynamically process user text based on OpenAI's GPT-4 model, interprets relevant instructions, and select from a predefined set of tools to fulfill tasks. The AI agents repeatedly cycle through a thought-action-observation process based on user input and selected tool output until it is determine that the final answer is obtained based on instructional prompt (examples in **Figures 2, S8, S9**).

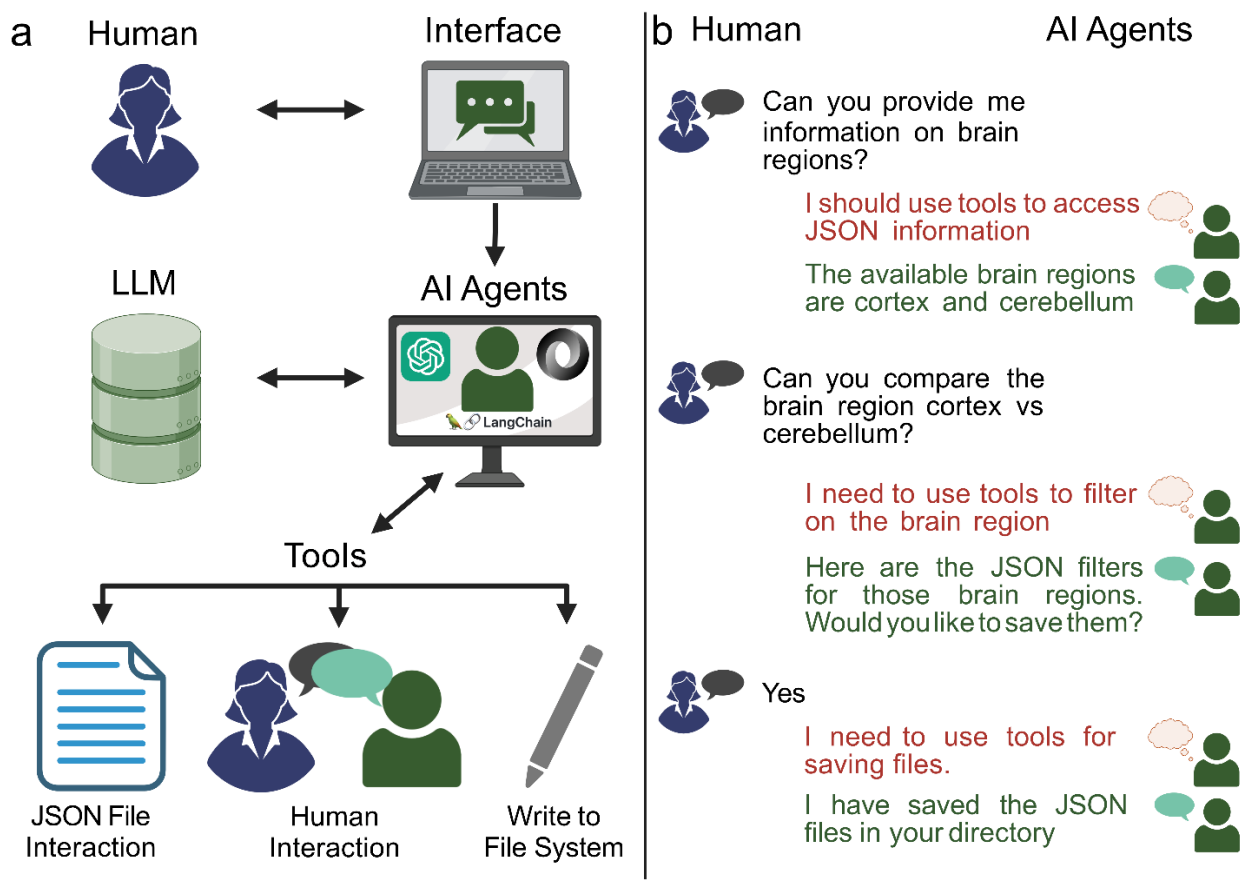

**Figure 2. Language User Interface (LUI) with chatbot and interactive artificial intelligent (AI) agents for lipidomics data analysis.** (a) A human uses CLAW's LUI to interact via a chatbot with customized AI agents for data manipulation. These AI agents use OpenAI's GPT-4 large language model (LLM) to process user requests on dataset filtration and then select from a list of tools developed for lipidomics analyses to take action to fulfill those requests. The custom prompt, used by AI agents for direction, instructs the agent to use tools in a guided manner, seeking user clarification and confirmation on requests to ensure robustness. (b) A simplified example of a human interacting with the AI agent via a chatbot interface. The human seeks information on available brain regions and then requests a comparison on the cortex vs cerebellum data. The AI agent thinks about the appropriate tools to fulfill the requests (red text) and responds in a conversational manner (green text).

The LangChain Python package was used and modified to create custom AI agents that parses and saves JSON files specific to dataset labeling and filtration. Specifically, LangChain's toolkit was extended, and a custom prompt was created to enable the agent to interact with the relevant JSON data in a highly structured manner. The toolkit included LangChain's built in JSON interaction tools, LangChain's tools for human input, a custom tool for filtering JSON files based on user input, and a modified version of LangChain's file writing tool to save JSON files that serve as CLAW statistical filters (**Figures S10-S12**). The agent's instructional prompt was customized to ensure robust operation by clearly defining goals, strictly outlining tool usage, and ensuring clarification is requested, if needed (**Figure S13**). The main goal of the agent is to write and save JSON filters for the current MRM dataset, with a subtask of answering user questions about the dataset. Each tool that the AI agent accesses contains its own section in the prompt that outlines conditions on when and how the tool can be used. Additionally, wording is added to instruct the agent to ask the user for clarification, whenever needed. This is done to make the AI agents behave in a robust and consistent manner. Specific instructions in the prompt allows the agent to generate two JSON files consecutively for comparison, facilitating potential enhancements for interactive data analysis. In **Video S1**, the LUI generates two JSON files to compare lipid MRM data from samples for 5xFAD vs wild-type genotypes for the cortex brain region of mice. In **Video S2**, even with deliberate user spelling errors and ambiguous queries, the LUI accurately identifies user's intent and generates the desired JSON files for analysis.

Generally, scientific data analysis tools require some level of programming experience to properly process relevant data. The LUI addresses this problem by assisting the user with CLAW data analysis. The LUI functions like a copilot or coscientist, performing user-defined queries and clarifying specifics regarding data analysis in an interactive manner. This enables users to issue commands via chat-based input, guiding the LUI agent to filter data for comparative analysis. To ensure robust performance, the LUI will request clarification when provided with invalid or unclear instructions (**Figure S8-S9**). We achieved this robust response by integrating native LLM with a specialized agent to utilize a select set of tools. To our knowledge, this represents the first use of an LLM to aid in lipidomics profiling, establishing a novel, robust and comprehensive approach for data analysis.

*EdgeR Statistical Analysis and Lipid Pathway Analysis*

Ion count measurements from MRM-profiling experiments follow a negative binomial distribution in a similar manner to data from RNA-seq experiments.[13,14] **Figure 3a** illustrates a representative distribution, while **Figure S17** showcases the true ion count distributions across all samples. Considering these distributions, tools such as edgeR's generalized linear model (GLM) developed for RNA-seq count data can be used for analysis of lipid MRM ion counts. By accounting for the inherent variability and structure in ion counts data, the EdgeR GLM approach ensures statistically robust calculation of changes in lipid abundance while incorporating experiment-specific variations in the blank (injection medium) for several injections. The edgeR package fits a GLM to the log–linear relationship for the mean variance as follows:

$$\log \mu_{ls} = X_s^T \beta_l + \log N_s$$

where $\mu_{ls}$ is the expected ion count for a particular lipid $l$ in a sample $s$ and $N_s$ represents the sum of all ion intensity for sample $s$. $\beta_l$ represents the regression coefficients associated with each lipid $l$ that captures the experimental conditions on lipid expression. The coefficient of variance (CV) is then calculated for lipid ion count for a given sample $y_{ls}$ using the following formula:

$$CV^2(y_{ls}) = 1/\mu_{ls} + \Phi_l$$

Ion count dispersion was listed as $\Phi_l$ and when multiple biological replicates are present, it is calculated using the common dispersion method.[53] When no biological replicates are present the dispersion term is set to 0.1 which is recommended for genetically identical model organisms.[54] Fold change is calculated between the groups of interest and p-values are obtained through the likelihood ratio test. The p-values are corrected using the Benjamini-Hochberg procedure to calculate false discovery rate (FDR). We considered FDR value of less than 0.1 (10%) as significant to identify differential lipids between comparisons.

Predicting the relationship between lipid molecules and the genes that encode proteins related to their biosynthetic pathways can help identify new targets for novel biological insights. Bioinformatics Methodology for Pathway Analysis (BioPAN) is a web-based tool that allows is users to upload lipidomic data for pathway analysis.[55] CLAW exports lipid MRM results as CSV files in a BioPAN-compatible format for each comparison selected. Instead of manually formatting lipidomics data, the CSV files generated by CLAW can be directly uploaded to the BioPAN website for downstream analysis.

**Results and Discussion**

*Brain Region-specific MRM profiling of Lipid Droplets*

CLAW was used to evaluate the differential expression of lipids within lipid droplets (LDs) isolated from specific brain regions (hippocampus, cortex, cerebellum, and diencephalon) of 18–24 month-old Alzheimer's disease model (5xFAD) and age-matched wild-type (WT) male mice. LDs are dynamic cellular organelles that not only serve as lipid reservoirs but play significant roles in cellular signaling, detoxification and inflammation.[56–61] Accordingly, their accumulation is correlated to pathophysiology of lipid imbalance linked disorders like Alzheimer's disease and even aging.[17,56,62–64] Despite being pointed out as "adipose saccules" in postmortem brain of AD patients by Alois Alzheimer's in 1907, functional significance of LDs has been largely overlooked.[65] Since dyshomeostatic environment in disease conditions manifest as aberrations in lipid droplet composition, we are elucidating the lipid signatures to find druggable targets for restoring brain homeostasis in AD and aging.[66–68]

CLAW's comparative analysis GUI (**Figure S7**) was used to rapidly select multiple comparisons from various brain regions from the 5xFAD and wild-type mice. This streamlined the analysis of MRM experimental data which was obtained using an Agilent 6495C mass spectrometer. We used tailored MRMs that were categorized into 10 main classes, including glycerophospholipids, glycerolipids, sphingolipids, fatty acyls (FA), and sterol lipids. Such broad coverage and depth of profiling enabled us to first identify a detailed LD composition. In brief, LDs were found to be rich in both TGs and CEs as traditionally characterized, but also contained a variety of lipid species spanning the acyl carnitine (CAR), sphingomyelin (SM), phosphatidylethanolamine (PE), and ceramide (CER) subclasses. For example, CAR 14:2, CAR 18:4, SM(d16:1/24:1), PE 38:0, and Cer(d18:1/16:0) were found across all LDs. Importantly, our bioinformatic analysis also facilitated the development of lipid-profile signature for LDs from aged-WT and 5xFAD brains, revealing that FA 24:6 was found to be downregulated in all 5xFAD samples. The combination of all brain regions resulted in 10 differentially expressed lipids between 5xFAD and WT samples. Specifically, FA 24:6, FA 24:5, and PG 32:5 down regulated, while 22:3 campestral ester, 22:2 campestral ester, 22:0 cholesteryl ester, and 22:1 cholesteryl ester were significantly upregulated lipids in 5xFAD mice.

Next, we constructed region-specific lipidomic profiles for LDs isolated from the brains of aged-WT and 5xFAD mice. The presence of cholesteryl ester-rich LDs in hippocampus and cortex, which are the hotspots for amyloid plaques, underscores the impact of environmental changes on composition in different brain regions.[17,69,70] Specifically, PG 32:5, PS 32:5, and FA 24:6 were down regulated in the hippocampus LDs of 5xFAD mice. In the diencephalon LDs from 5xFAD mice, FA 24:6 was also downregulated, while PS 32:5, 22:3 campestral ester, 22:1 campestral ester, 22:2 campestral ester, 22:0 cholesteryl ester, and 22:1 cholesteryl ester was found to be upregulated in LDs. In the cerebellum, there was only a single differentially expressed lipid, fatty acid 24:6 when comparing LDs from 5xFAD to WT mice. **Figure 3 d-f** illustrates the log-fold change (logFC) of lipids in 5xFAD mice relative to wild type across various lipid classes, highlighting the differential expression that varies depending on the brain region. In the hippocampus, acyl carnitines, cholesterol esters, phosphatidylserine (PS), and triacylglycerols are upregulated in the 5xFAD model. While, phosphatidylcholine (PC), phosphatidylinositol (PI), phosphatidylglycerol (PG), and sphingomyelin (SM) classes are downregulated in this region for the 5xFAD mice. In the diencephalon, there is an upregulation of acyl carnitine, cholesterol esters, phosphatidylcholine, and sphingomyelin lipid classes. However, certain fatty acids exhibit downregulation in the diencephalon of the 5xFAD model. For the cerebellum, the lipid classes of acyl carnitines, ceramides, and phosphatidylserine are downregulated, while phosphatidylinositol, sphingomyelin, and phosphatidylcholine classes exhibit upregulation. In the cortex, specific cholesterol esters and phosphatidylglycerol classes are downregulated, with phosphatidylinositol showing an upregulation. Notably, acyl carnitines, phosphatidylcholine, and sphingomyelin display bimodal distributions, with lipids from these classes being both up and downregulated. Intriguingly, fatty acids displayed the most expansive distribution across all brain regions, with instances of both upregulation and downregulation. The differential lipidomic profiles across various brain regions underscore the importance of understanding the regional differences in lipid metabolism resulting from neurodegenerative disorders. The consistent downregulation of fatty acid 24:6, regardless of the brain region, suggests it may play a pivotal role in the pathophysiology of 5xFAD mice, warranting further investigation into its potential role. Overall, these results demonstrate CLAW's ability to manage complex lipidomic datasets, allowing efficient and meaningful comparisons that could shed light on the role of lipid droplets in the pathogenesis of neurodegenerative diseases. Significant lipids and their corresponding LogFC are shown in **Table S2-S6**. All MRM transitions utilized in this study, along with their corresponding lipid classes and abbreviations, are detailed in **Table S1.** A visual representation of the MRM transition count distribution is shown in **Figure 3b**. PCA was used to visualize variance among the samples. In **Figure 3c**, a PCA was constructed by summing the intensities across all brain regions for each sample. The analysis revealed marginally greater variance among the wildtype samples compared to the 5xFAD. This observation aligns with the heatmap generated using the same summation approach in **Figure 3d**. Heatmaps of individual brain regions comparing 5xFAD vs WT are shown in **Figure S18. Figure S19** displays PCAs for 5xFAD vs. WT lipids specifically within the cerebellum, cortex, diencephalon, and hippocampus. However, these PCAs do not present any discernible patterns, making it challenging to draw definitive conclusions. This lack of clear patterns aligns with the high variance seen in the data. **Figure S20** illustrates the total ion count intensities for the monitored reaction monitoring (MRMs) of lipids investigated in this study, with LDs from 5xFAD mice exhibiting higher lipid intensities across all regions combined, suggesting an increased presence of lipids. Conversely, **Figure S21** provides a region-specific analysis, confirming the overall trend; however, it notably reveals that in the hippocampus, wild-type (WT) samples display greater lipid intensity, deviating from the pattern observed in other regions. Uniform Manifold Approximation and Projection (UMAP) was used to depict the variability in lipid classes across different brain regions for both 5xFAD and WT genotypes shown in **Figure S22**. The UMAP made with all brain regions combined shows greater variability among all lipid classes in 5xFAD compared to WT (**Figure S23**). CLAW automatically exports a variety of plots, including PCAs, to assist users in interpreting their data. A detailed guide to running CLAW is provided in **Supporting Information.**

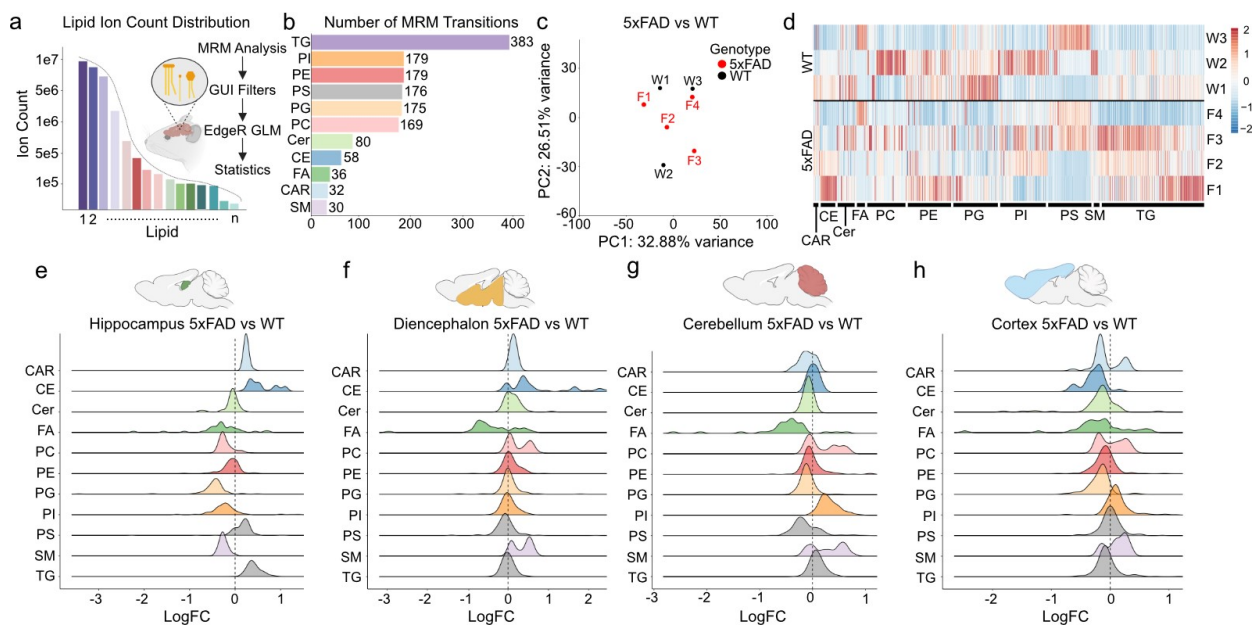

**Figure 3. Comparison of brain region resolved LD composition for 5xFAD vs WT male mice.** a) Experimental design illustrates the processing of a typical negative binomial distribution of lipid MRM counts, employing edgeR's GLM for statistical analysis of significant lipids b) Distribution of MRM transitions selected for screening lipids. A total of 1497 transitions (used to ID lipid species) were organized into 10 MRM-based mass spectrometry methods for lipid classes. c) PCA demonstrating the variation in the LD lipidome of four brain regions. d) Heatmap that displays z-scores of normalized lipid intensities across samples. e) Ridge plot displaying distribution of logFC values for all lipid species within each class in Hippocampus f) Diencephalon g) Cerebellum and h) Cortex region of 5xFAD vs WT mice brain. Data represented for LDs obtained from hippocampus, diencephalon, cerebellum, and cortex region isolated from 5xFAD (n=5 mice) and WT mice (n=3) brain.

### Identification of TGs in Canola Oil with C=C specificity

CLAW was used to investigate the lipid profiles of canola oil across three stages in the refinement process. The LC-OzESI-MRM experiments were performed on an Agilent 6495C QQQ modified for OzESI as described in Randolph et al.[51] Triacylglycerols (TGs) containing an unsaturated fatty acyl chain FA 18:1 (i.e., 18 carbons, 1 C=C) were targeted due to their known high abundance in canola oil.[71,72] Due to their known biological prevalence, we sought to determine the relative ratio of TGs containing FA 18:1 with an unsaturation site either Δ9 or Δ7 carbons away from the terminal, methyl end of the aliphatic chain. Note that we denote the first isomer as TG_FA 18:1(*n*-9) and the second as TG_FA 18:1(*n*-7). To evaluate these double bond positions with the CLAW OzESI method, the user selected these positions shown in **Figure 4a** using a Jupyter Notebook. Briefly, LC was used to first separate TG molecular species based on their equivalent carbon number (ECN), as highlighted with the total ion chromatogram (TIC) shown in **Figure 4c**. In general, the larger the ECN, the greater the retention time. For example, the most abundant TG molecular species, TG 54:3 and TG 52:2, both characterized by an ENC of 48, eluted at 18.1 min. TG 54:2

(ECN = 50) had a retention time of 20.0 min, while TG 52:3 and TG 54:4 (ECNs = 46) coeluted at 16.1 min. **Figure 4c** depicts the results of "ozone on" and "ozone off" experiments. Briefly, the black and red traces depict the TIC for "ozone off" and "ozone on" experiments, respectively. When the ozone gas is supplied to the nebulizer, the unsaturated TGs readily react with gaseous ozone, resulting in a decrease in the observed TIC signal.

Utilizing tabulated ozonolysis neutral loss values that have been extensively reported in the literature, we first tabulated a list of predicted TG precursor ion values ([TG + NH4]$^+$) and the corresponding ozonolysis product ions.$^{42,73–75}$ This tabulation occurs automatically by a python script in **Figure 4b** allowing CLAW to rapidly generate possible m/z values for each user selected double bond location for TGs in canola oil. For example, the ammonium cation adduct of TG 54:4 is observed at *m/z* 900.8. To monitor TG 54:4 molecular species containing the FA 18:1, we established an MRM precursor/product ion pair of *m/z* 900.8 → 601.6 by exploiting the NL of 299.2 Da indicative of FA 18:1. Based on predicted NL values following ozonolysis for monounsaturated lipids, TG 54:4 that carries an *n*-9 and *n*-7 double bond generated the diagnostic product ions at *m/z* 790.6 and *m/z* 818.7, respectively. Next, exploiting the NL of FA 18:1, OzESI-MRMs can be established to monitor unsaturated lipid profiles with C=C specificity. Thus, to monitor the presence of TG 54:4_FA 18:1*n*-9 the precursor/product ion pair of *m/z* 790.6 → 690.9 was used, while the precursor/product ion pair of *m/z* 818.7 → 690.9 was used to profile TG 54:4_FA 18:1*n*-7. To demonstrate the LC-OzESI-MRM approach, **Figures 4d** and **4e** reflect the EICs of the above MRM transitions for TG 54:4 with ozone off and on, respectively. Briefly, when the ozone is off, only signal for the TG 54:4_FA 18:1 MRM transition (i.e., m/z 900.8 → 601.6) is observed. However, when ozone is admitted to the nebulizer, additional MRM signal is observed for the n-9 and n-7 C=C isomer channels for TG 54:4_FA 18:1. **Figure 4f** presents lipid ratio between the *n*-9 and *n*-7 isomers of a variety of TG molecular species in canola oil containing FA 18:1. In general, the isomeric ratio of TGs containing FAs 18:1-*n*-9 and *n*-7 remained relatively consistent in canola oil through its refinement stages. Thus, we can conclude that the refinement process does not significantly impact isomer ratios. While we were not surprised by consistent isomer ratios displayed throughout the refinement process, it is important to note that lipid precursor ion populations previously grouped as a single entity, in fact are composed of at least two distinct isomeric populations, previously unresolved with conventional LC-MS/MS approaches. To our knowledge, this is the first critical investigation of TG profiles at the C=C level in canola oil throughout the refinement stages.

To validate CLAW isomer ratio calculation results, CLAW outcomes were cross-referenced with the manual LC peak area data analysis (**Tables S7-S8** and **Figures S14-S16**). Traditionally, relative abundance calculations such as the isomer ratios portrayed herein, can be achieved via exploiting integrated peak area values generated within commercial data-processing software. However, peak area calculations are most reliable when analyte baseline separation can be achieved, requiring highly optimized and precise LC methods. Thus, we chose to employ maximum MRM intensity values as a basis for CLAW's isomer ratio calculations as an alternate strategy. Notably, both CLAW and manual analysis showed consistent agreement. For example, an overall average standard deviation of 0.75 was observed between these two methods across all oil samples. Highlighting the precision and reliability of the CLAW's data processing capabilities, CLAW TG isomer ratio values calculated using maximum MRM intensity values revealed an average standard deviation of 0.29 for seven lipids across the three canola oil purities. In contrast, a higher average standard deviation of 0.37 was observed for TG isomer ratio values obtained via manual LC peak area calculations. Moreso, standards deviations of 0.84, 0.92, and 0.50 for crude, degummed, and RBD canola oils, respectively, were observed across automated CLAW and manual isomer ratio calculations.

The observed discrepancies can be ascribed to CLAW selecting the maximum MRM intensity value from the raw data, while manual interpretation is subject to variations in peak area selection. Future work aims to further investigate the discrepancies arising between isomer ratio calculations that rely on LC peak area versus maximum MRM intensity values.

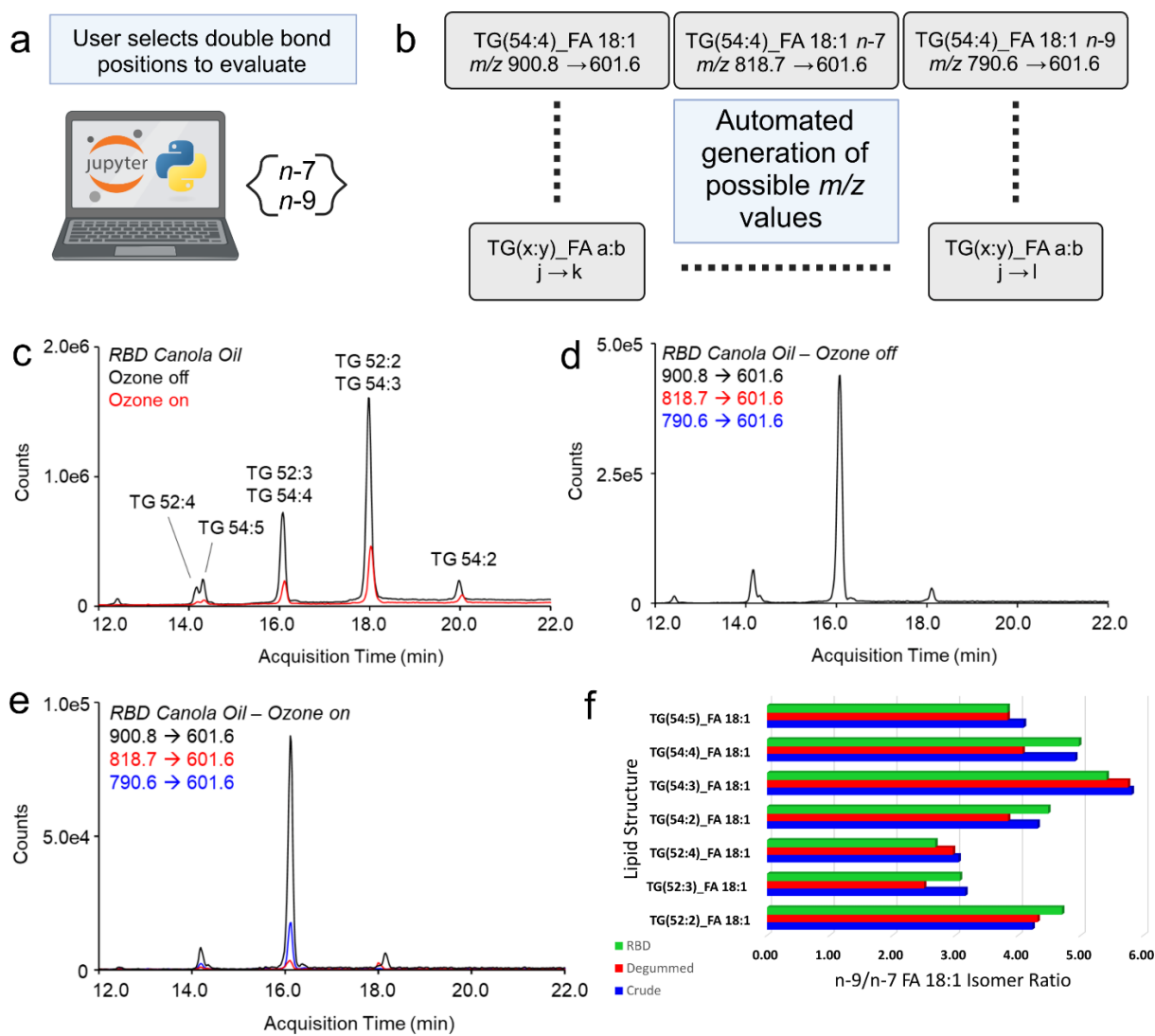

**Figure 4. CLAW OzESI MRM methodology.** a) User selects n-7 and n-9 double bond positions to evaluate b) Possible m/z values are automatically generated based on the user selected double bond positions and then matched and annotated with the experimental data c) Comparison of ozone off/on chromatogram for RBD canola oil with TG retention time identified d) RBD canola oil chromatogram with ozone off showing MRM transition of TG(54:4) FA 18:1, but not the corresponding double bond locations e) RBD canola oil chromatogram with ozone on showing MRM transition of TG(54:4) FA 18:1 and the corresponding n-9 and n-7 double bond position f) Results of the n-9/n-7 FA 18:1 max intensity isomer ratio in TGs for each processing step of canola oil.

## Conclusions

CLAW has been developed as a comprehensive MRM-based pipeline for the detailed identification of lipids in complex biological samples. By automating processes traditionally done manually, CLAW significantly reduces analysis time and boosts throughput. In large-scale lipidomics, experimental worklist generation is tedious and susceptible to human-related errors which CLAW provides a solution by facilitating the generation of acquisition worklists. Following MRM-based experiments, raw data is converted to an mzML file format, which is then parsed by a CLAW python script using the pymzml package. During parsing, the data is annotated by matching the experimental results to a user selected MRM database. The annotated results are stored in a flexible pandas dataframe to easily extract the necessary data for statistical and bioinformatic analysis. An important statistical analysis feature in CLAW is the integration of edgeR GLMs, marking a significant advancement in lipidomics data analysis by addressing the inherent variability of lipid MRM ion count distributions with enhanced reliability in differential expression analysis. While CLAW represents a comprehensive solution, we advocate in general for the broader adoption of statistical methodologies, like edgeR's GLM, to effectively handle over dispersed ion counts in MS-based lipidomics. To appeal to a broad range of users, CLAW's GUI/LUI simplifies complex dataset analysis by enabling simultaneous comparison of multiple parameters. To improve user experience, the LUI provides a chatbot-style interaction that aids in analysis and data processing.

To demonstrate the capabilities of CLAW, we initially used the developed pipeline on biological samples from four distinct mouse brain regions. In the first example, traditional MRM-profiling of roughly 1500 individual lipid species was conducted on LDs isolated from specific brain regions of 5xFAD and WT mice. The data shows clear lipidome distinctions amongst LDs obtained from aged and AD-diseased brains, indicating the LDs related to aging and AD are not the same. Furthermore, distinct lipid signatures for the LDs isolated from the hippocampus, cortex, cerebellum, and diencephalon regions were observed.

In the second example, an online LC-OzESI-MRM method previously developed in-house was used to examine TGs with C=C specificity from several samples of canola oil taken at various stages of refinement. TGs profiled using OzESI-MRMs revealed little to no effect of the refinement process of TG isomer composition. While not surprising, the utilization of OzESI-MRMs in conjugation with CLAW successfully resolved, identified, and relatively quantified isomeric populations of TG molecular species that would otherwise remain unidentified using conventional or traditional workflows.

Future work aims to extend CLAW's worklist generator to other acquisition software beyond MassHunter. Specifically, the parsing and annotation will be extended beyond raw data files from Agilent instruments to other types of mass spectrometers. In addition, the integration of AI agents as co-pilots and agent-to-agent communications will be developed for robust and responsible planning and execution of automated experiments utilizing LLMs and tools framework outlined in this work. We also aim to enhance our use of edgeR methodology by incorporating multi-factor experimental designs and evaluating both trended and tagwise dispersion estimates, as opposed to solely using common dispersion, and compare the likelihood ratio test with the quasi-likelihood (QL) F-test. Software support will also be developed for other types of instrument modifications. Specifically, in addition to OzESI module, new modules are being developed to support lipid identification combined with chemical conjugation techniques.

**Code Availability**

https://github.com/chopralab/CLAW.git

**Acknowledgements**

This work was supported, in part, by the NSF I/UCRC Center for Bioanalytical Metrology (Award 1916991), the United States Department of Defense USAMRAA award W81XWH2010665 through the Peer Reviewed Alzheimer's Research Program, the National Institutes of Health (NIH) award, R01MH128866 by National Institute of Mental Health, and NIH National Center for Advancing Translational Sciences U18TR004146 and ASPIRE Challenge and Reduction-to-Practice awards and Defense Threat Reduction Agency (DTRA) AIMS-HITS contract award MCDC2202-003 to G.C. The Purdue University Center for Cancer Research funded by NIH grant P30 CA023168 is also acknowledged. The authors would like to thank Agilent Technologies Inc. for the donation of the Triple Quadrupole LC/MS instrument to Chopra Laboratory. The authors thank Dr. Stephen J. Blanksby and Dr. Berwyck Poad, along with other members of the Blanksby lab, for the generous loan of necessary equipment and assistance to set-up ozone electrospray ionization mass spectrometry (OzESI-MS) capabilities within the Chopra laboratory. The authors thank all members of the Center for Bioanalytical Metrology for their guidance and support. We thank Dr. Kanchan Bisht for helping with isolating different sections of the brain. The content is solely the responsibility of the authors and does not necessarily represent the official views of the National Institutes of Health. CER acknowledges funding from the Arnold O. Beckman Postdoctoral Fellowship program.

**Competing Interests**

The authors declare the following competing financial interest(s): G.C. is the Director of the Merck-Purdue Center funded by Merck Sharp & Dohme, a subsidiary of Merck and the co-founder of Meditati Inc and BrainGnosis Inc. M.M is a co-founder of Meditati Inc. The remaining authors declare no competing interests.

# Supporting Information

**Comprehensive Lipidomic Automation Workflow using Large Language Models**

Connor Beveridge[1,#], Sanjay Iyer[1,#], Caitlin E. Randolph[1,#], Matthew Muhoberac[1], Palak Manchanda[1], Amy C. Clingenpeel[2], Shane Tichy[3], Gaurav Chopra[1,4]*

[1]Department of Chemistry, Purdue University, West Lafayette, IN, 47907, [2]ExxonMobil Technology and Engineering Company, Annandale, NJ, 08801; [3]Agilent Technologies Inc. Santa Clara, CA 95051; [4]Department of Computer Science (*by courtesy*), Purdue University, West Lafayette, IN, 47907

[#]These authors contributed equally to this work

*Corresponding Author: gchopra@purdue.edu

## Table of Contents

# CLAW Workflow Example

```python
#Import all the necessary libraries
import pymzml
import csv
import os
import pandas as pd
import numpy as np
import math
from matplotlib import pyplot as plt
import re
import plotly.express as px
from collections import defaultdict

import plotly.io as pio
import json
import plotly.graph_objs as go
import matplotlib.colors as mcolors

import json
import ipywidgets as widgets
from IPython.display import display

import warnings
from IPython.display import display, Image, clear_output
import time

##Custom SCripts

from NO_AVERAGE_SCRIPTS import full_parse
from NO_AVERAGE_SCRIPTS import filter_dataframe
from NO_AVERAGE_SCRIPTS import hex_to_rgba_hex
from NO_AVERAGE_SCRIPTS import json_to_string
from NO_AVERAGE_SCRIPTS import prep_edge_R

from NO_AVERAGE_SCRIPTS import make_pie_chart_no_replicates

from NO_AVERAGE_SCRIPTS import average_pie_chart_no_repeats

from NO_AVERAGE_SCRIPTS import make_bar_plot_comparisons
import os
from IPython.display import display
from ipywidgets import widgets
# dfdfplotting_functions

Pre_folder = './Projects/'

from tools.GUI import assign_blank, load_blank_name
from tools.GUI import folder_navigator, load_project_folder

from tools.GUI import filter_samples, load_data_labels
from tools.GUI import display_pair_widgets, load_data, remove_empty_entries, get_unique_json_objects

from tools.parsing import add_suffix
```

1. Execute the Jupyter cell to import all required packages.

```python
file_name_to_save = '5xFAD_vs_WT_Brain_LD'
extra_name = ""

tolerance = 0.1

save_data= True
load_previously_parsed = False
custom_data=True
```

2. Fill out the essential variables:

a. file_name_to_save: Designate the name for the file to be saved. If *load_previously_parsed* is set to true, this will be the name of the file to be loaded.
b. extra_name: Add an extension to filenames for repeated experiments or to monitor samples removed during filtering.
c. Tolerance: Set the *m/z* matching threshold for transitions. For example, a value of 0.1 will match transitions within that range.
d. save_data: Determine if the data should be saved for subsequent analysis. This will use the *file_name_to_save* to determine the file name.
e. load_previously_parsed: Opt to load data previously parsed. This will use the *file_name_to_save* to determine the file to be loaded.
f. custom_data: Decide between using the default MRM database or a customized one. If choosing a custom database, you must upload a *Custom_MRMs.csv* file.

```
folder_navigator()
Project_Folder = load_project_folder()
```

[Navigate]

[Select this folder]

[Select Current Folder]

```
..
BRAIN_5xFAD_LD_CLAW_Manuscript
NEW_5xFAD_LIVER_BRAIN_LD
Burda_Lab
LIVER_LD_5XFAD
BRAIN_LIVER_LD
ELOVL1
Brain_5xFAD
```

```
Contents of /home/cbeveri/lipid_parser2/Lipidomics/lipid_platform/Projects:
BRAIN_5xFAD_LD_CLAW_Manuscript
NEW_5xFAD_LIVER_BRAIN_LD
Burda_Lab
LIVER_LD_5XFAD
BRAIN_LIVER_LD
ELOVL1
Brain_5xFAD
```

3. GUI to navigate to current project folder. The project folder should be selected and set up as described in the read me. Select the button "Navigate" to move directories. "Select this Folder" selects the highlighted folder. "Select Current Folder" selects the current directory.

Samples [Blank ▾]

[Assign Blank]

4. Use the GUI to select the "Blank" based on Sample Names from the "labels.csv" file.

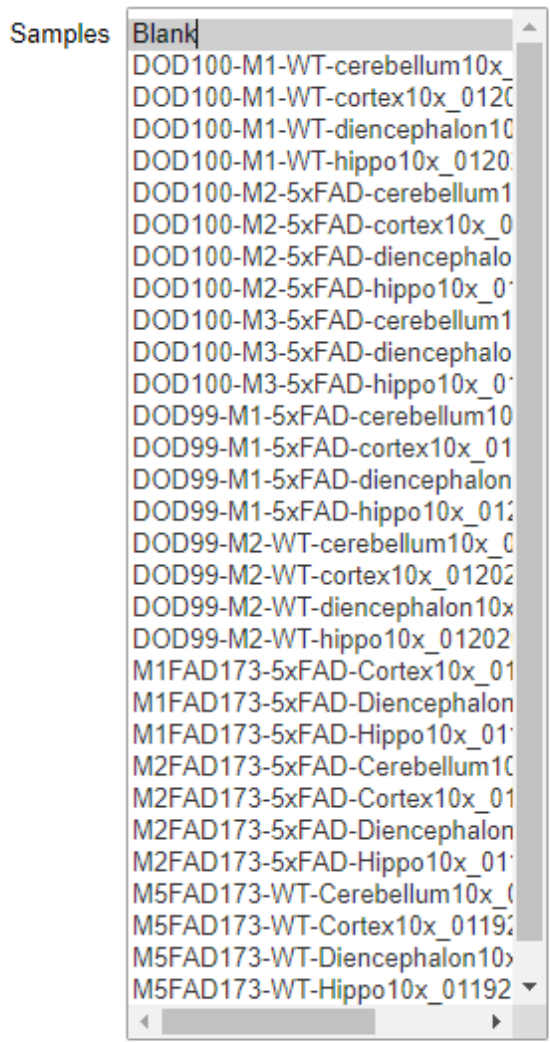

5. Use the sample filtering GUI to highlight any undesired samples and click "filter samples" to exclude them from the analysis, allowing for selective removal of specific samples.

[Screenshot of GUI showing Genotype (WT, 5xFAD), Brain Region (cerebellum, cortex, diencephalon, hippo), and Sex (Male) selection boxes in two columns, with "Finish" and "Add more JSON pairs" buttons.]

6. Example GUI for selecting comparison analysis. In the example above it is Genotype: 5xFAD Brain Region: cerebellum Sex: male in the left column and Genotype: WT Brain Region: cerebellum Sex: male in the right column. This will perform comparative analysis between these two groups where an upregulated lipid is up in the 5xFAD mouse (chosen via the left column). A user can create an additional filter for comparative analysis by clicking the "Add more JSON pairs" or if a user is finished the "Finish" button can be selected.

```python
# Ignore warnings
warnings.filterwarnings('ignore')

# Display GIF
gif = Image(filename='Figures/cat_gif.gif')
display(gif)

print("Your data is PURRing...")

if load_previously_parsed == True:
    df_matched = pd.read_csv(Project_Folder+"Processed Results/"+file_name_to_save+".csv")

else:
    df_matched = full_parse(data_base_name_location,mzml_folder, folder_name_to_save,labels_df, blank_name,
                            file_name_to_save,tolerance, custom_data = custom_data, remove_std = remove_std,save_data=save_data)

json_list_pairs = load_data()
json_list_pairs = remove_empty_entries(json_list_pairs)
json_list_singles = get_unique_json_objects(json_list_pairs)
make_pie_chart_no_replicates(df_matched,plots_2_save_path,json_list_singles,labels_list,blank_name,extra_name)
average_pie_chart_no_repeats(df_matched,plots_2_save_path,json_list_singles,labels_list,blank_name,extra_name)
make_bar_plot_comparisons(df_matched, plots_2_save_path, json_list_pairs,labels_list,blank_name,extra_name)
labels_list = labels_list + ['method_type',"Transition"]
df_matched = add_suffix(df_matched)
combined_df = prep_edge_R(df_matched,json_list_pairs,Pre_edge_r_path,blank_name,labels_list,extra_name)
!bash myjob.sh
```

7. Upon utilizing the GUI/LUI to establish JSON filters, execute the final cell in the Jupyter notebook.

## Automatic Worklist Generator

The CLAW workflow initiates with an automatic worklist generator that requires three CSV files: "Path_info.csv," "labels.csv," and "methods.csv."

In labels.csv (illustrated in **Figure S1**), the columns 'Sample Name' and 'Position' are mandatory. Additional columns like Cage, Sex, Genotype, and Brain Region can be incorporated to describe samples. These labels facilitate data filtering in the GUI/LUI, as depicted in **Figure S7**.

Path_info.csv (displayed in **Figure S2**) contains three essential paths:
1. path_to_methods indicates where the method directory location.
2. path_to_save_data which indicates the storage location for saved files.
3. Clean_file_method, specifying the path to the desired method file for cleaning.

The third CSV, "methods.csv" (shown in **Figure S3**), contains an 'annotation' column with abbreviations for the Mass Hunter datafile. The "method name" column identifies the method file to utilize, which is merged with the method path from "Path_info.csv."

To run the automatic worklist generator run the script use the jupyter notebook worklist_generator.ipynb
1. Ensure the required CSV files are in the same directory as the script.
2. Execute the script. During its run:
   - Input the desired date and run details. This data will be incorporated into the saved file name for streamlined experiment tracking.
   - Specify the number of technical replicates, representing repeated injections of the same sample. The minimum requirement is one replicate.
   - If cleaning between injections is desired:
     - Define after how many sample injections cleaning is to be performed.
     - Indicate the total number of cleaning cycles.
     - Decide if cleaning is required between different methods.
   - If no cleaning is desired, simply input "0" for cleaning intervals.

This design streamlines the worklist generation process, saving time and reducing the potential for manual errors.

A sample completed worklist is provided in **Figure S4**. The columns represent sample position, the method and its path, and the save file with its path, respectively. This format allows easy copying to Mass Hunter, reducing redundancy. An illustration of the overall workflow is shown in **Figure S5**.

# SUPPORTING FIGURES

**Figure S1. Example of labels.csv file**. The automatic worklist generator uses this file to denote sample name and positions while the additional columns assist in filtering the data. Sample Name and Position columns are mandatory to fill out for the automatic worklist generator to function. Also, each instance in the Sample Name column must be unique. Additional columns can also be for data filtering with the Graphical User Interface (GUI) or Language User Interface (LUI). In the example above Cage, Sex, Genotype, and Brain Region information has also been added to each sample and thus the GUI/LUI can filter using that information.

**Figure S2. Illustration of a completed *Path_info.csv* file**. This file is only required for the automatic worklist generator. Three essential paths need to be specified: 1. Path to methods: The directory where method files are located for Agilent Mass Hunter software (.m). 2. path_to_save_data: The destination directory for saving output files generated by Agilent Mass Hunter software (.d). 3. Clean_file_method: The full path to the cleaning method file to be used after a user designated number of injections. These paths should be specific to the users own system.

| | A | B |
|---|---|---|
| 1 | annotation | method name |
| 2 | FFA | MRM profiling_FA_05-10-2022.m |
| 3 | TAG14_0 | MRM profiling_TAG14_0_05-10-2022.m |
| 4 | TAG16_0 | MRM profiling_TAG16_0_05-10-2022.m |
| 5 | TAG16_1 | MRM profiling_TAG16_1_05-10-2022.m |
| 6 | TAG18_0 | MRM profiling_TAG18_0_05-10-2022.m |
| 7 | TAG18_1 | MRM profiling_TAG18_1_05-10-2022.m |
| 8 | TAG18_2 | MRM profiling_TAG18_2_05-10-2022.m |
| 9 | TAG18_3 | MRM profiling_TAG18_3_05-10-2022.m |
| 10 | TAG20_4 | MRM profiling_TAG20_4_05-10-2022.m |
| 11 | TAG22_5 | MRM profiling_TAG22_5_05-10-2022.m |
| 12 | TAG22_6 | MRM profiling_TAG22_6_05-10-2022.m |
| 13 | PG | MRM profiling_PG_07-08-2022.m |
| 14 | PCandSM | MRM profiling_LPC_PC-O_PC-P_PC_05-04-2022.m |
| 15 | PE | MRM profiling_LPE_PE-O_PE-P_PE_05-17-2022.m |
| 16 | PI | MRM profiling_PI_07-08-2022.m |
| 17 | PS | MRM profiling_PS_07-08-2022.m |
| 18 | DAG16_0 | MRM profiling_DAG16_0_06-22-2022.m |
| 19 | DAG16_1 | MRM profiling_DAG16_1_06-22-2022.m |
| 20 | DAG18_0 | MRM profiling_DAG18_0_06-22-2022.m |
| 21 | DAG18_1 | MRM profiling_DAG18_1_06-22-2022.m |
| 22 | DAG18_2 | MRM profiling_DAG18_2_06-22-2022.m |
| 23 | CE | MRM profiling_cholesteryl esters_07-08-2022.m |
| 24 | CER | MRM profiling_Ceramides_09-13-2022.m |
| 25 | AC | MRM profiling_Acylcarnitines_05-05-2022.m |

**Figure S3. Showing example of *methods.csv* file**. This file denotes what Agilent Mass Hunter MRM methods will be used during analysis. The annotation column represents the abbreviation used when saving the Mass Hunter datafile. The method name column lists the method file to be used. This method will integrate with the path specified in *Path_info.csv*. Ensure these methods correspond precisely with those in your Agilent Mass Hunter.

| | A | B | C |
|---|---|---|---|
| 1 | P1-A1 | D:\MassHunter\methods\MRMs\MRM profiling_FA_05-10-2022.m | D:\MassHunter\Data\Brain_5xFAD_LD\FFA_DOD100-M1-WT-cerebellum_BrainLD_07_20_2023.d |

**Figure S4. Partial example of generated worklist**. This displays a partial example of a finalized worklist generated with the automatic worklist generator. The first column indicates the sample position, the second showcases the method with its respective path, and the third highlights the save file and its path. Each column is designed to be copy-pasted into Mass Hunter Software. After filling out the three columns the user can then begin data acquisition using Mass Hunter.

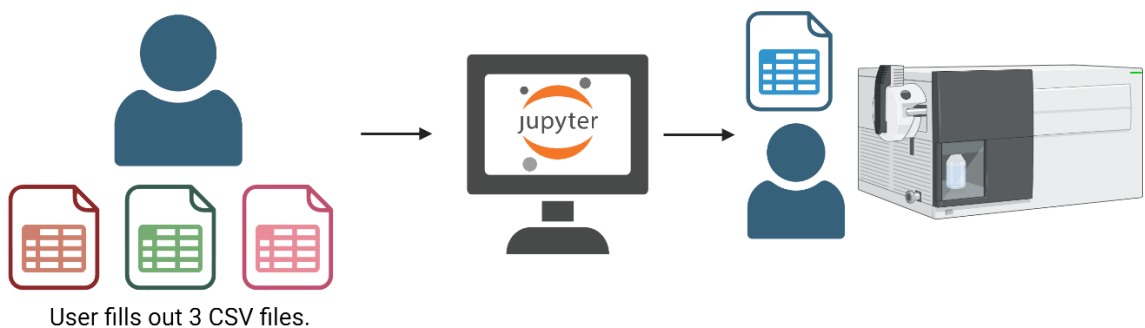

User fills out 3 CSV files.

**Figure S5. Worklist Generator Workflow.** The user will fill out 3 CSV files containing pathing information, methods information, and sample information as described in **Figures S1-S3**. A new CSV file will be generated, detailing the sample position, the associated method file, and the destination for saving results. An example of this is shown in **Figure S4.**

| | A | B | C | D |
|---|---|---|---|---|
| 1 | Lipid | Class | Precursor_Ion | Product_Ion |
| 2 | TAG(50:4)_FA 18:2 | TAG | 846.8 | 549.8 |
| 3 | TAG(52:0)_FA 18:2 | TAG | 880.8 | 583.8 |
| 4 | TAG(52:0)_FA 20:0 | TAG | 880.8 | 551.8 |
| 5 | TAG(52:0)_FA 20:4 | TAG | 880.8 | 559.8 |
| 6 | TAG(52:1)_FA 18:2 | TAG | 878.8 | 581.8 |
| 7 | TAG(52:1)_FA 20:4 | TAG | 878.8 | 557.8 |
| 8 | TAG(52:1)_FA20:0 | TAG | 878.8 | 549.8 |
| 9 | TAG(52:2)_FA 18:2 | TAG | 876.8 | 579.8 |
| 10 | TAG(52:2)_FA 20:0 | TAG | 876.8 | 547.8 |
| 11 | TAG(52:2)_FA 20:4 | TAG | 876.8 | 555.8 |
| 12 | TAG(52:3)_FA 18:2 | TAG | 874.8 | 577.8 |
| 13 | TAG(52:3)_FA 20:0 | TAG | 874.8 | 545.8 |
| 14 | TAG(52:3)_FA 20:4 | TAG | 874.8 | 553.8 |
| 15 | TAG(52:4)_FA 18:2 | TAG | 872.8 | 575.8 |
| 16 | TAG(52:4)_FA 20:0 | TAG | 872.8 | 543.8 |
| 17 | TAG(52:4)_FA 20:4 | TAG | 872.8 | 551.8 |
| 18 | TAG(52:5)_FA 18:2 | TAG | 870.8 | 573.8 |

**Figure S6. Example of custom MRM database**. The user is given an option to upload their own database containing MRM transitions. The user must fill out lipid names, class, precursor ion $m/z$, and production ion $m/z$. All columns are required to be filled out. CLAW uses this information to parse and annotate the .mzml files generated through mass spectrometry lipidomics experiments. Any transitions observed in the .mzml files that aren't listed here will be excluded during the parsing process.

**Figure S7. Example of CLAW's interactive GUI.** CLAW's GUI allows quick comparison of multiple groups. The groups available are based on the *labels.csv* file. For instance, the left column showcases Genotype: 5xFAD, Brain Region: cerebellum, Sex: male, while the right presents Genotype: WT, Brain Region: cerebellum, Sex: male. The analysis will compare these groups, highlighting lipids that are upregulated in the 5xFAD mouse from the left column. Users can incorporate additional filters for the comparison by selecting the "Add more JSON pairs" button. Once all selections are made, the "Finish" button finalizes the process.

```
> Entering new AgentExecutor chain...
Thought: I should use the json_spec_list_keys tool to see what keys exist in the JSON
Action: json_spec_list_keys
Action Input: data
Observation: ['Cage', 'Sex', 'Genotype', 'Brain Region']
Thought: I should ask the human what key and values they would like to filter on
Action: Human
Action Input: What key and values would you like to filter on?

What key and values would you like to filter on?
What are the brain option and Gene options?

Observation: What are the brain option and Gene options?
Thought: I should use the json_spec_get_value tool to see what values exist for the Brain Region and Genotype keys
Action: json_spec_get_value
Action Input: data["Brain Region"]
Observation: ['hippocampus', 'cortex', 'cerebellum', 'diencephalon']
Thought: I should use the json_spec_get_value tool to see what values exist for the Brain Region and Genotype keys
Action: json_spec_get_value
Action Input: data["Genotype"]
Observation: ['WT', '5xFAD']
Thought: I should use the json_filter_json_from_lists tool to filter the JSON based on the keys and values provided by the human
Action: json_filter_json_from_lists
Action Input: ["Brain Region","Genotype"];["hippocampus","WT"]
Observation: {'Genotype': ['WT'], 'Brain Region': ['hippocampus']}
Thought: I should use the Human tool to ask the human to confirm that this is the JSON they wanted to filter
Action: Human
Action Input: Is this the JSON you wanted to filter?

Is this the JSON you wanted to filter?
No I want 5xFAD and cortex

Observation: No I want 5xFAD and cortex
Thought: I should use the json_filter_json_from_lists tool to filter the JSON based on the keys and values provided by the human
Action: json_filter_json_from_lists
Action Input: ["Brain Region","Genotype"];["cortex","5xFAD"]
Observation: {'Genotype': ['5xFAD'], 'Brain Region': ['cortex']}
Thought: I should use the Human tool to ask the human to confirm that this is the JSON they wanted to filter
Action: Human
Action Input: Is this the JSON you wanted to filter?

Is this the JSON you wanted to filter?
yes

Observation: yes
Thought: I should use the write_file tool to save the JSON
Action: write_file
Action Input: {
    "path": "./demo_data/JSON/filter_keys_1.json",
    "text": {'Genotype': ['5xFAD'], 'Brain Region': ['cortex']}
}
Observation: File written successfully to ./demo_data/JSON/filter_keys_1.json.
```

**Figure S8. Example of Language User Interface (LUI) interaction**. The user asks, "What are the brain option and gene options?" The LUI then infers that the user would like to know the potential labels for both genotype and brain region. The LUI then creates a filter for WT and hippocampus based on user input and will then ask for clarification. The user specifies the following parameters: 5xFAD and cortex. The LUI then creates the correct JSON filter but will seek user confirmation before proceeding. After confirmation the LUI will then save the JSON.

```
Observation: File written successfully to ./demo_data/JSON/filter_keys_1.json.
Thought: I should ask the human what key and values they would like to filter on
Action: Human
Action Input: What key and values would you like to filter on?

What key and values would you like to filter on?
wild type from the cortex

Observation: wild type from the cortex
Thought: I should use the json_filter_json_from_lists tool to filter the JSON based on the keys and values provided by the huma
n
Action: json_filter_json_from_lists
Action Input: ["Brain Region","Genotype"];["cortex","WT"]
Observation: {'Genotype': ['WT'], 'Brain Region': ['cortex']}
Thought: I should use the Human tool to ask the human to confirm that this is the JSON they wanted to filter
Action: Human
Action Input: Is this the JSON you wanted to filter?

Is this the JSON you wanted to filter?
yes

Observation: yes
Thought: I should use the write_file tool to save the JSON
Action: write_file
Action Input: {
    "path": "./demo_data/JSON/filter_keys_2.json",
    "text": {'Genotype': ['WT'], 'Brain Region': ['cortex']}
}
Observation: File written successfully to ./demo_data/JSON/filter_keys_2.json.
Thought: I now know the final answer
Final Answer: Meow! I have successfully saved two filtered JSONs based on the original JSON. You can find them in the demo_dat
a/JSON folder.

> Finished chain.

'Meow! I have successfully saved two filtered JSONs based on the original JSON. You can find them in the demo_data/JSON folde
r.'
```

**Figure S9**. **Example of LUI interaction continued.** After creation of the first JSON file the LUI will then ask for the parameters of the second JSON file. The user responds with "wild type from the cortex." Although the label for genotype is "WT" the LUI can infer the meaning of wild type and can recognize the cortex is a brain region. The LUI will once again confirm before creating the second JSON file.

```python
Class JsonFilterFromKeysTool(BaseTool):

    def filter_json_nested(self, json_dict, keys_string, values_string):
        # Parse keys_string and values_string into lists
        keys = ast.literal_eval(keys_string)
        values = ast.literal_eval(values_string)

        filtered_dict = {}
        for k, v in json_dict.items():
            if k in keys:
                # If value is a list, filter the list based on the allowed values
                if isinstance(v, list):
                    filtered_list = [item for item in v if item in values]
                    if not filtered_list:
                        return f"Error: The list for key {k} is empty after filtering."
                    Filtered_dict[k] = filtered_list
                elif isinstance(v, dict):1.
```

```
                    Filtered_dict[k] = filter_json_nested(v, keys_string,
values_string)
                elif v in values:  # For non-list, non-dict values
                    filtered_dict[k] = v
        return filtered_dict

    name="json_filter_json_from_lists"
    description = """"Can be used to filter a JSON at a provided path based on
provided keys and values.
        Before calling this you should be SURE that the path to this exists.
        The input is an exact text representation of two Python lists 13eparated
by a semi colon where first list is the keys  to filter on, and the second list
is the values to filter on.
        An example is as follows '["key1","key2"];["value1","value2"]'
    """"
    spec: JsonSpec

    def _run(
        self,
        tool_input: str,
        run_manager: Optional[CallbackManagerForToolRun] = None,
    ) -> str:
        return self.filter_json_nested(
            self.spec.dict_,
            tool_input.split(";")[0],
            tool_input.split(";")[1]
        )

    async def _arun(
        self,
        tool_input: str,
        run_manager: Optional[AsyncCallbackManagerForToolRun] = None,
    ) -> str:
        return self._run(tool_input)
```

**Figure S10**. **Language User Interface (LUI) tool setup.** Custom LangChain agent tool for filtration of JSON files based on key and value information provided by the user. These filtered JSON files will be used as filters for data for statistical analysis. The tools private run function (called upon agent use) is assiged to return the output of the custom filtration function (`filter_json_nested`) provided with a list of keys and values. The custom filtration function recursively searches through the JSON file and only pulls enteries which match the key and values provided. If provided keys and/or values are not present, the function will return an error string which will be caught by the agent and the agent will attemp to find keys and values which are valid and similar to user input or ask the user for clarification. Once a sutiable filtration of the original JSON has been obtained, the filter will be provided to the agent for future use.

```python
class CustomWriteFileTool(BaseFileToolMixin, BaseTool):

    name: str = "write_file"
    # args_schema: Type[BaseModel] = WriteFileInput
    description: str = "Write file to disk"

    def _run(
        self,
        info: str,
        append: bool = False,
        run_manager: Optional[CallbackManagerForToolRun] = None,
    ) -> str:
        file_path = str(ast.literal_eval(info)["path"])
        text = str(ast.literal_eval(info)["text"])
        try:
            write_path = self.get_relative_path(file_path)
        except FileValidationError:
            return INVALID_PATH_TEMPLATE.format(arg_name="file_path", value=file_path)
        try:
            write_path.parent.mkdir(exist_ok=True, parents=False)
            mode = "a" if append else "w"
            with write_path.open(mode, encoding="utf-8") as f:
                f.write(text)
            return f"File written successfully to {file_path}."
        except Exception as e:
            return "Error: " + str(e)

    async def _arun(
        self,
        file_path: str,
        text: str,
        append: bool = False,
        run_manager: Optional[AsyncCallbackManagerForToolRun] = None,
    ) -> str:
        # TODO: Add aiofiles method
        raise NotImplementedError
```

**Figure S11**. **LUI tool setup continued**. Modified LangChain agent file writing tool for writing a JSON filter to disk. LangChain base file writing tool was modifed to be specifically compatible with the JSON filters which we use. This tool attempts to write a file to disk at the provided path with the provided text. In CLAW usage, the text will be JSON filter information that was the successful output of the custom JSON filtration tool.

```python
Class CustomToolkit(BaseToolkit):

    spec: JsonSpec

    def get_tools(self) -> List[BaseTool]:
        return [
            JsonListKeysTool(spec=self.spec),
            JsonGetValueTool(spec=self.spec),
            JsonFilterFromKeysTool(spec=self.spec),
            HumanInputRun(),
            CustomWriteFileTool()
        ]
```

**Figure S12**. **LUI custom toolkit setup**. Custom toolkit provided to the AI agent which is used in CLAW's language user interface (LUI). This toolkit contains LangChain's builtin tools for JSON file interaction (JSONListKeysTool and JSONGetValueTool), LangChain's builtin tool for humaninteraction (HumanInputRun), the custom tool for JSON file filtration, and the modified version of LangChain's write file tool.

```
prefix = """You are an agent designed to interact with a JSON and a human.

Your overall goal is to interact with the user to create and save TWO (2) new
filtered JSONs based on the original JSON.
During interaction with the human prior to your final answer, they may ask for
information about the keys or values of the JSON which you should provide to
them.
You should talk like a cat when responding, make sure to use lots of meows.

ONLY provide your final answer after you have successfully use the `write_file`
tool twice without an error. You MUST fulfill this prior to providing your final
answer.

You have access to the following tools which help you learn more about the JSON
you are interacting with and provide the human with information on the JSON you
are interacting with.
Only use the information returned by the below tools to construct your final
answer.

You MUST use the `json_filter_json_from_lists` tool in the following manner:

1. Your input to the `json_filter_json_from_lists` tool must be exactly two
Python formatted lists in the form '["key1","key2"];["value1","value2"]''.
```

the first list is a list of dictionary keys and the second list is a a list of
dictionary values. The input to the tool must have its list be seperated by a
semicolon and nothing else.
You can infer any keys from previous observations of the `json_spec_list_keys`
tool and you can infer any values from previous observations of the
`json_spec_get_value` tool.

2. If the observation from the `json_filter_json_from_lists` tool returns an
error, you must use the `json_spec_list_keys` and `json_spec_get_value` tools to
find valid keys and values for this tool and
then infer based on the humans request.

3. After using the `json_filter_json_from_lists` tool without an error, you MUST
use the `Human` tool with an input of a string of the filtered JSON and ask them
to confirm that it is the JSON they wanted to filter.

You MUST use the `write_file` tool in the following manner:

1. Before using the `write_file` tool, if you do not have a previous observation
from the `json_filter_json_from_lists` tool, use the `Human` tool with an input
asking they would like to filter on.

2. Your input for the `write_file` tool should be a text representation in JSON
format containing the keys "path" and "text".
The path should be "./demo_data/JSON/filter_keys_<n>.json" where <n> is a string
of the number 1 or 2. 1 if it is the first JSON that is being save, 2 if it is
the second JSON being saved
The text should be a JSON formatted string based on a previous observation from
the `json_filter_json_from_lists` tool.

3. Check to see how many times you have used the `write_file` tool without error,
if it is 2 or more times, you should return your final answer.

4. After using the write_file tool without error, use the `Human` tool with an
input asking what key and values they would like to filter on.

You MUST use the `json_spec_list_keys` and `json_spec_get_value` tools in the
following manner:

1. Your input to tools named `json_spec_list_keys` and `json_spec_get_value`
should be in the form of `data["key"]` where `data` is the JSON blob you are
interacting with, and the syntax used is Python.
You should only use keys that you know for a fact exist. You must validate that a
key exists by seeing it previously when calling `json_spec_list_keys`.
If you have not seen a key in one of those responses, you cannot use it.

```
You should only add one key at a time to the path. You cannot add multiple keys
at once.
If you encounter a "KeyError", go back to the previous key, look at the available
keys, and try again.

2. If the human asked you to provide them with keys or values of the JSON and you
used either the `json_spec_list_keys` or `json_spec_get_value` tool, you MUST use
the `Human` tool with an input of what you observed.

You MUST use the `Human` tool in the following manner:

1. Do not give the `Human` tool any input on keys or value that you did not
directly observe as a result of using the `json_spec_list_keys` or
`json_spec_get_value` tools

1. If you are providing the human with keys or values from the JSON, you must
first use the `json_spec_list_keys` or `json_spec_get_value` tools to confirm
those keys and values exist in the JSON.
You can use the `json_spec_list_keys` or `json_spec_get_value` tools any number
of times prior to using the `Human` tool.

2. If you just used another tool besides 'Human' in the previous action. The
input to the `Human` tool should include your observation of that action.

3. If you are asking the human for clarification or confirmation, you input the
the `Human` tool should be explicit on what you need clarification or
confirmation on.

If the question does not seem to be related to the JSON, use the `Human` tool and
ask for guidance
Always begin your interaction with the `json_spec_list_keys` tool with input
"data" to see what keys exist in the JSON.
Then for each key that you observe use the `json_spec_get_value` tool for that
key to see what values exists in the JSON.
Then start your interaction with the human.

Note that sometimes the value at a given path is large. In this case, you will
get an error "Value is a large dictionary, should explore its keys directly".
In this case, you should ALWAYS follow up by using the `json_spec_list_keys` tool
to see what keys exist at that path.
Do not simply refer the user to the JSON or a section of the JSON, as this is not
a valid answer. Keep digging until you find the answer and explicitly return it.
"""

suffix = """Begin!"
```

```
Remember to format your answers talking as a cat would.

{chat_history}
Question: {input}
{agent_scratchpad}"""
memory = ConversationBufferMemory(memory_key="chat_history")
```

**Figure S13. Example LUI prompt designed to instruct the LLM model**. Custom prompt designed to provide LUI instruction to the AI agent. The agent is provided with this prompt at object definition and references the prompt as it cycles through its thought-action-observation process. The begenning of the prompt defines the overall goal of the agent which is to filter up to two JSON based on user specifications with a subtasks of providing the user with informaiton which they request. The middle of prompt outlines correct usage of the tools and how to handel any erros that may result. The end of the prompt tells the agent how to start its interaction with the user and tells the agent to ask for guidance if needed. This strict structure and detailed wording of the prompt promotes robust agent interaction.

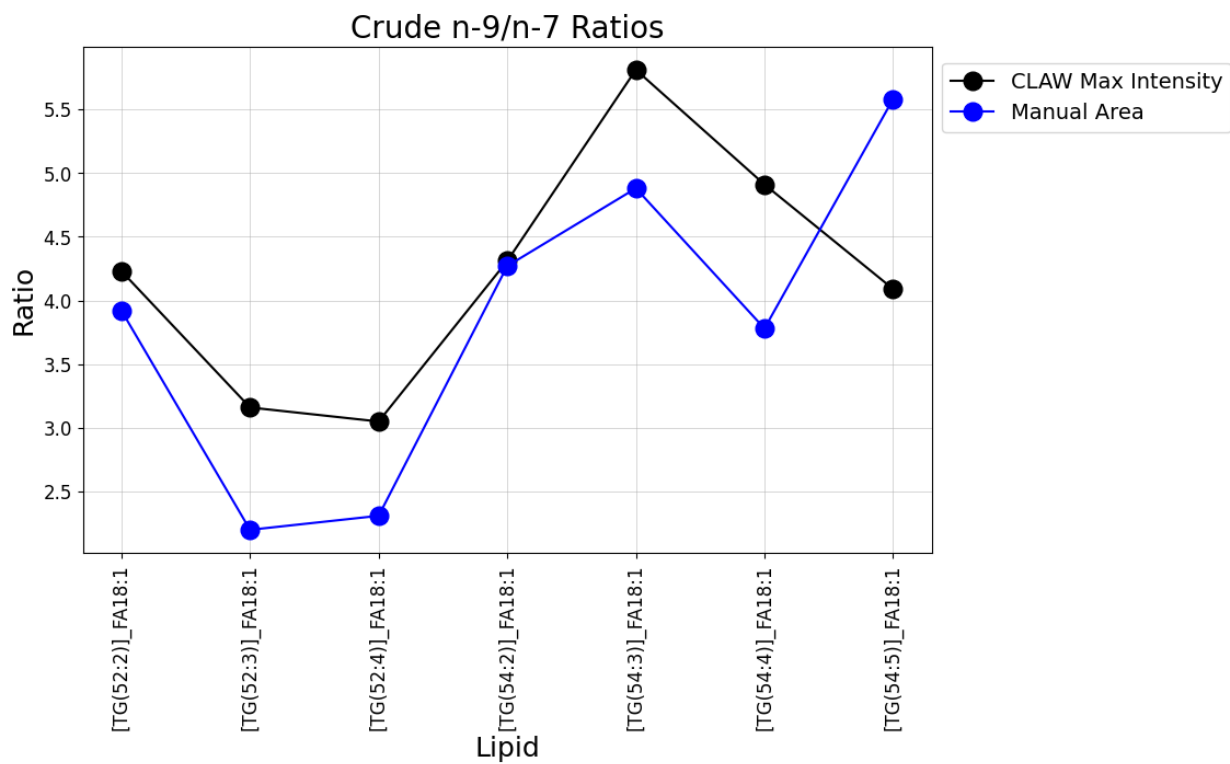

**Figure S14**. **Crude canola oil comparison of CLAW max intensity ratios vs manual area ratios.** Results of CLAW max intensity ratios and manually calculated area ratios for the OzESI crude canola oil data. The ratio of the n-9/n-7 double bond location was compared at seven different TGs.

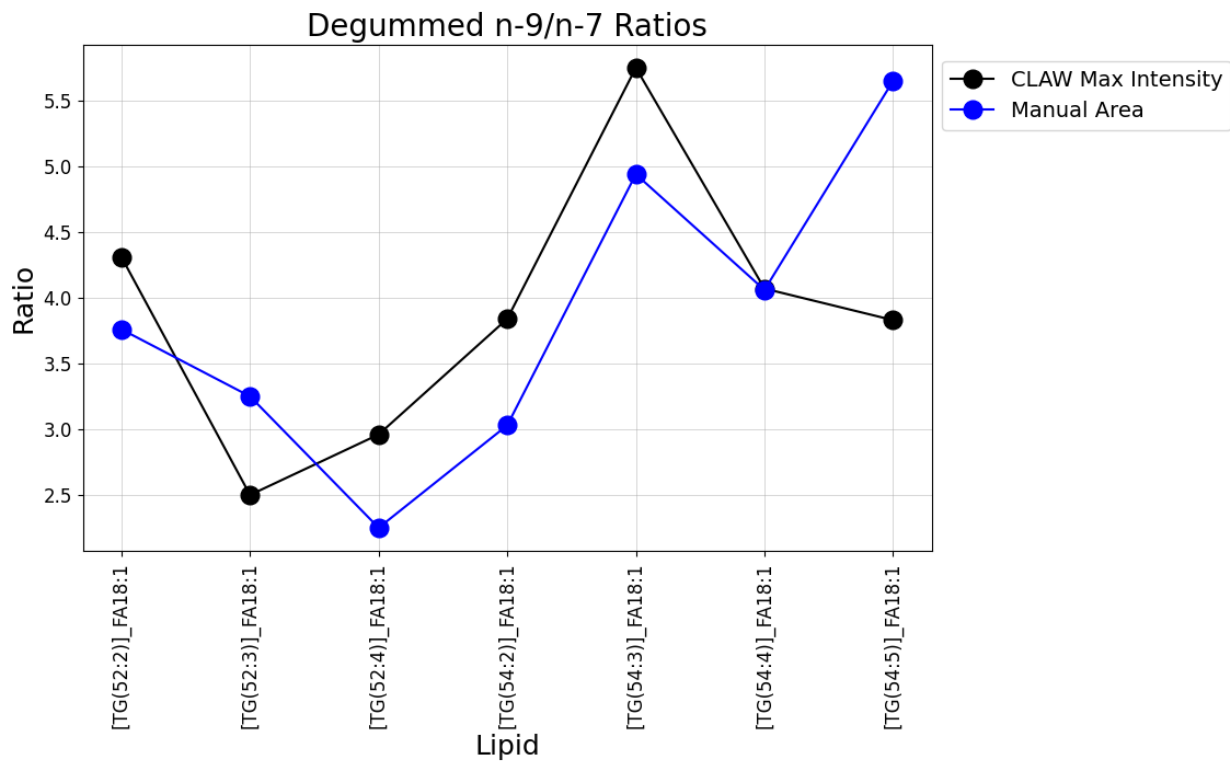

**Figure S15**. **Degummed canola oil comparison of CLAW max intensity ratios vs manual area ratios.** Results of CLAW max intensity ratios and manually calculated area ratios for the OzESI degummed canola oil data. The ratio of the n-9/n-7 double bond location was compared at seven different TGs.

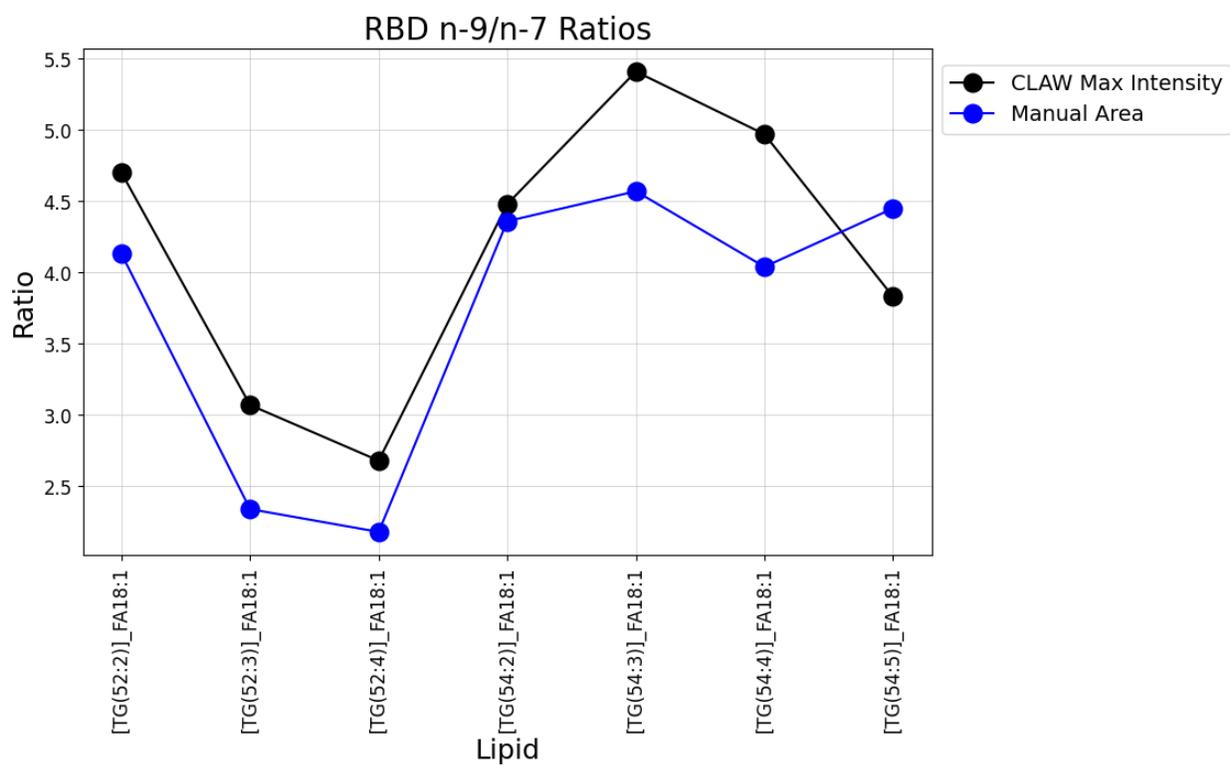

**Figure S16**. **RBD canola oil comparison of CLAW max intensity ratios vs manual area ratios.** Results of CLAW max intensity ratios and manually calculated area ratios for the OzESI RBD canola oil data. The ratio of the n-9/n-7 double bond location was compared at seven different TGs.

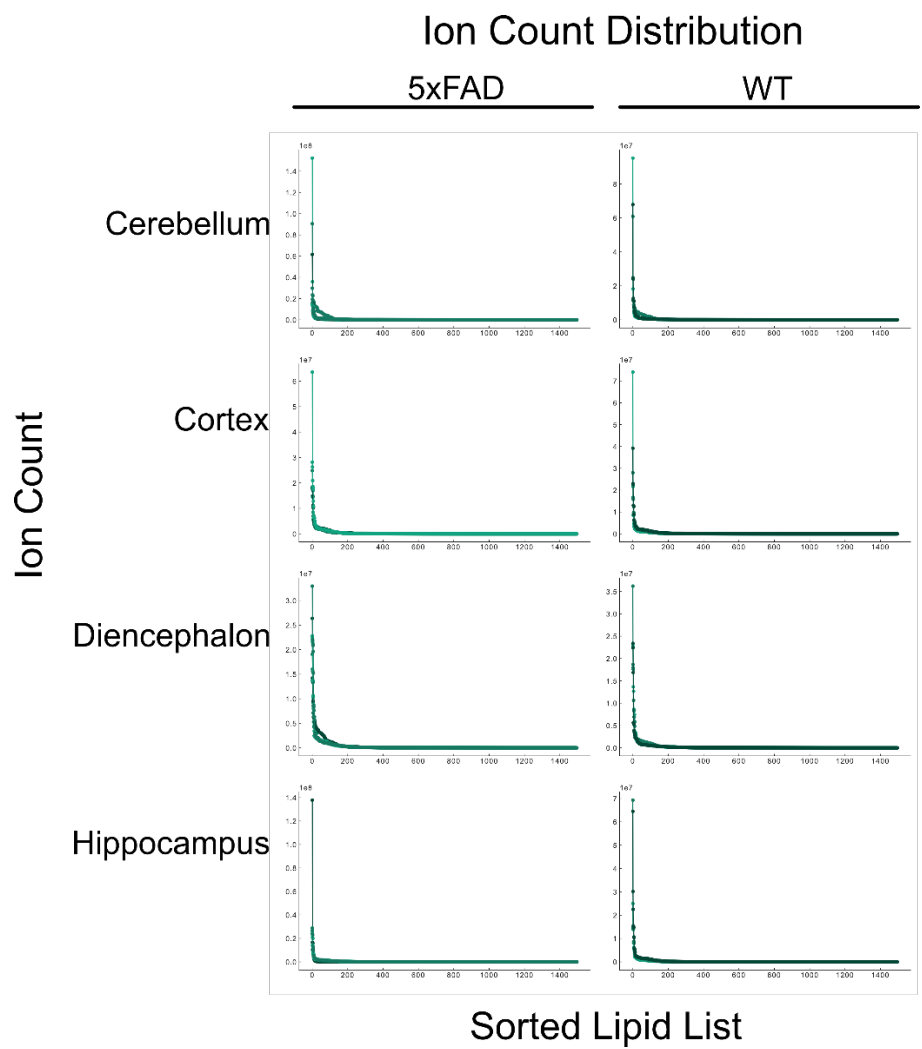

**Figure S17**. **Ion count distribution of all samples**. The ion count follows a negative binomial distribution for each sample. Statistical models used are tailored to this distribution.

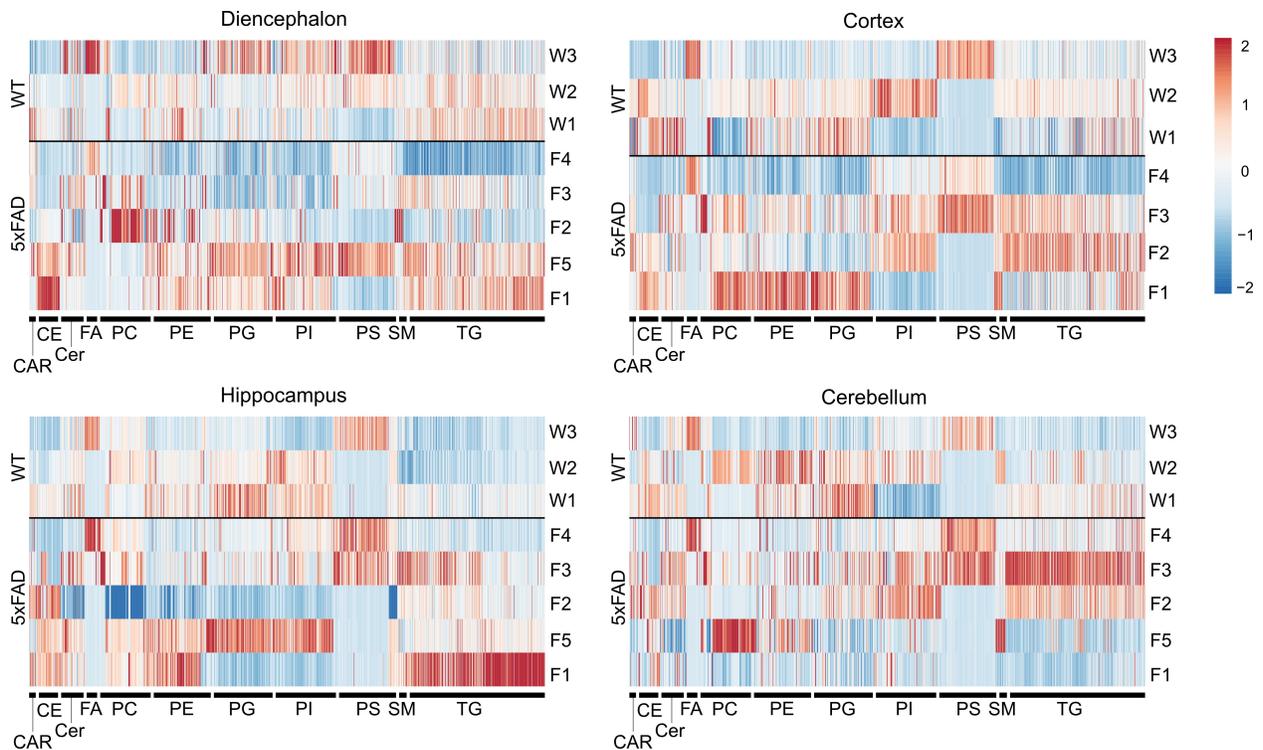

**Figure S18. Brain region-specific lipid class heatmaps for 5xFAD and WT mice.** The heatmap displays z-scores of lipid intensities across samples. Values were derived by subtracting blank measurements, with any resulting values below zero adjusted to zero. Each sample is normalized based on its total intensity, ensuring the sum of lipid intensities for any given sample is 1, before calculating the z-scores. Intensities from the brain region diencephalon, cortex, hippocampus, and cerebellum are shown separately. The color scale is based on z-scores computed row-wise, where each row denotes a specific lipid, highlighting variations in lipid abundance across samples. The heatmaps show high variation among both wildtype and 5xFAD mice as well as differences in lipid droplet profiles from different brain regions.

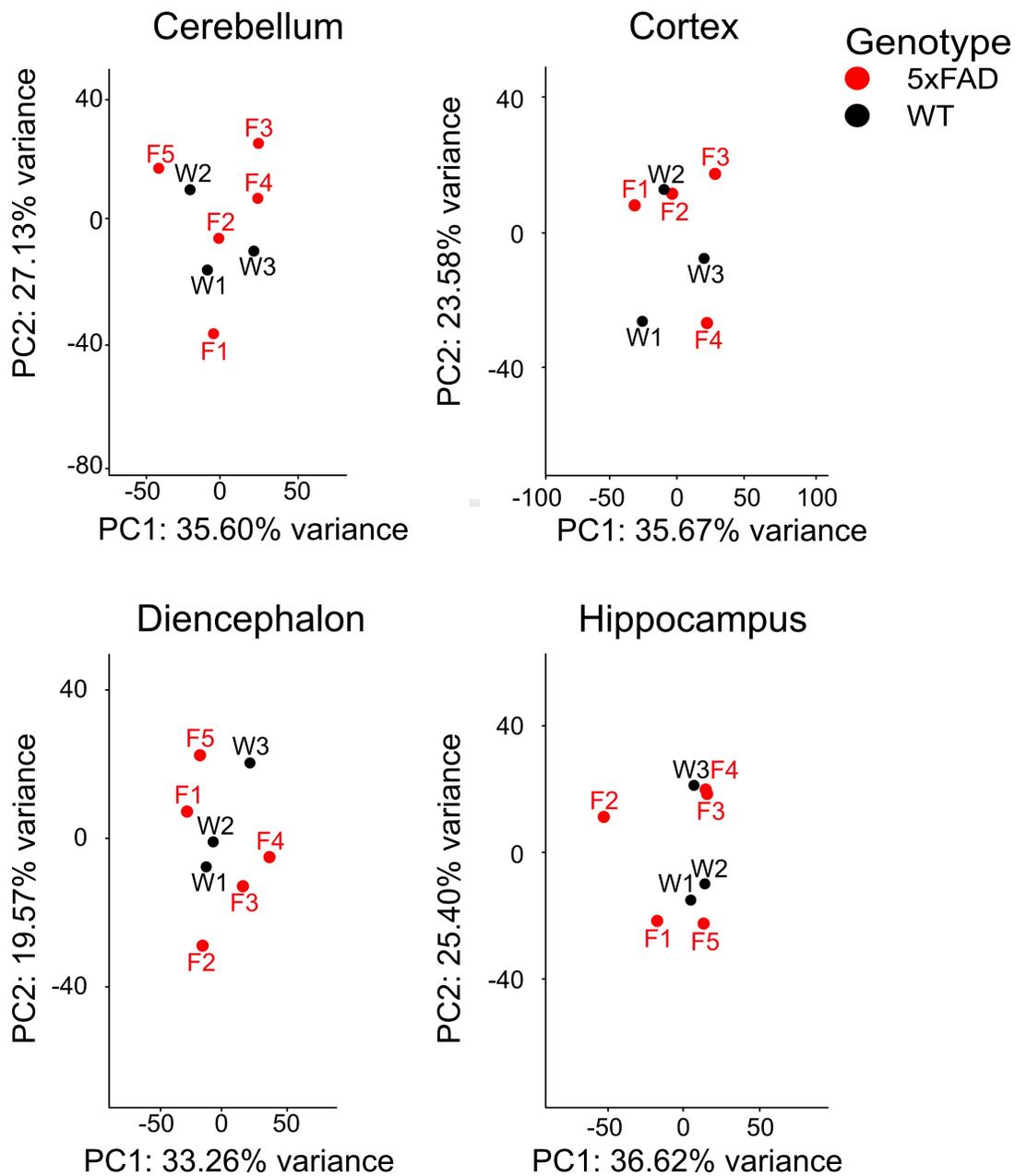

**Figure S19**. PCA plot of all lipids for 5xFAD vs WT in brain regions cerebellum, cortex, diencephalon, and hippocampus. High variance is shown in both 5xFAD and wildtype mice regardless of brain region.

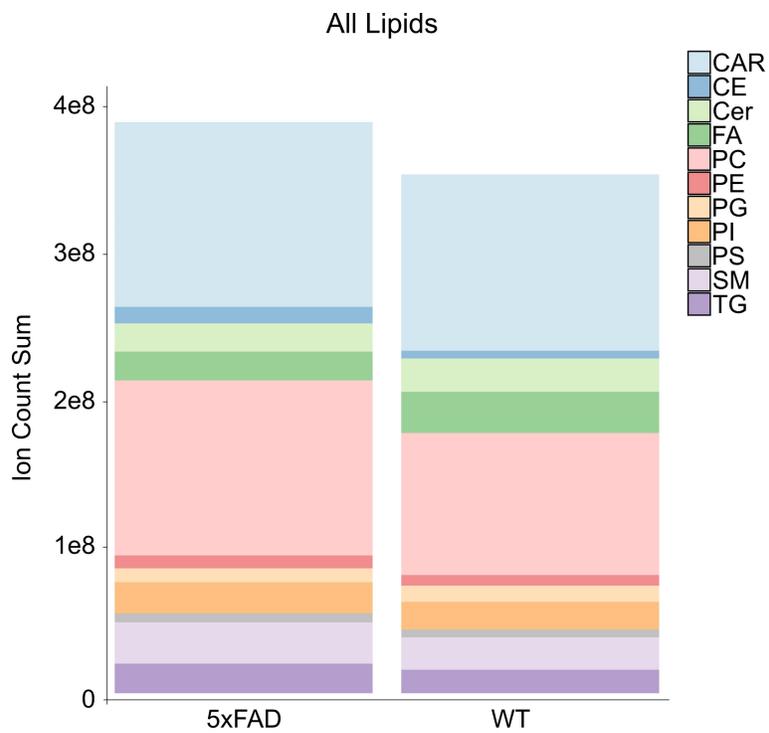

**Figure S20**. **Summed ion counts for all lipids in 5xFAD vs WT**. Showing greater lipid expression in lipid droplets from brains of 5xFAD relative to wild type mice.

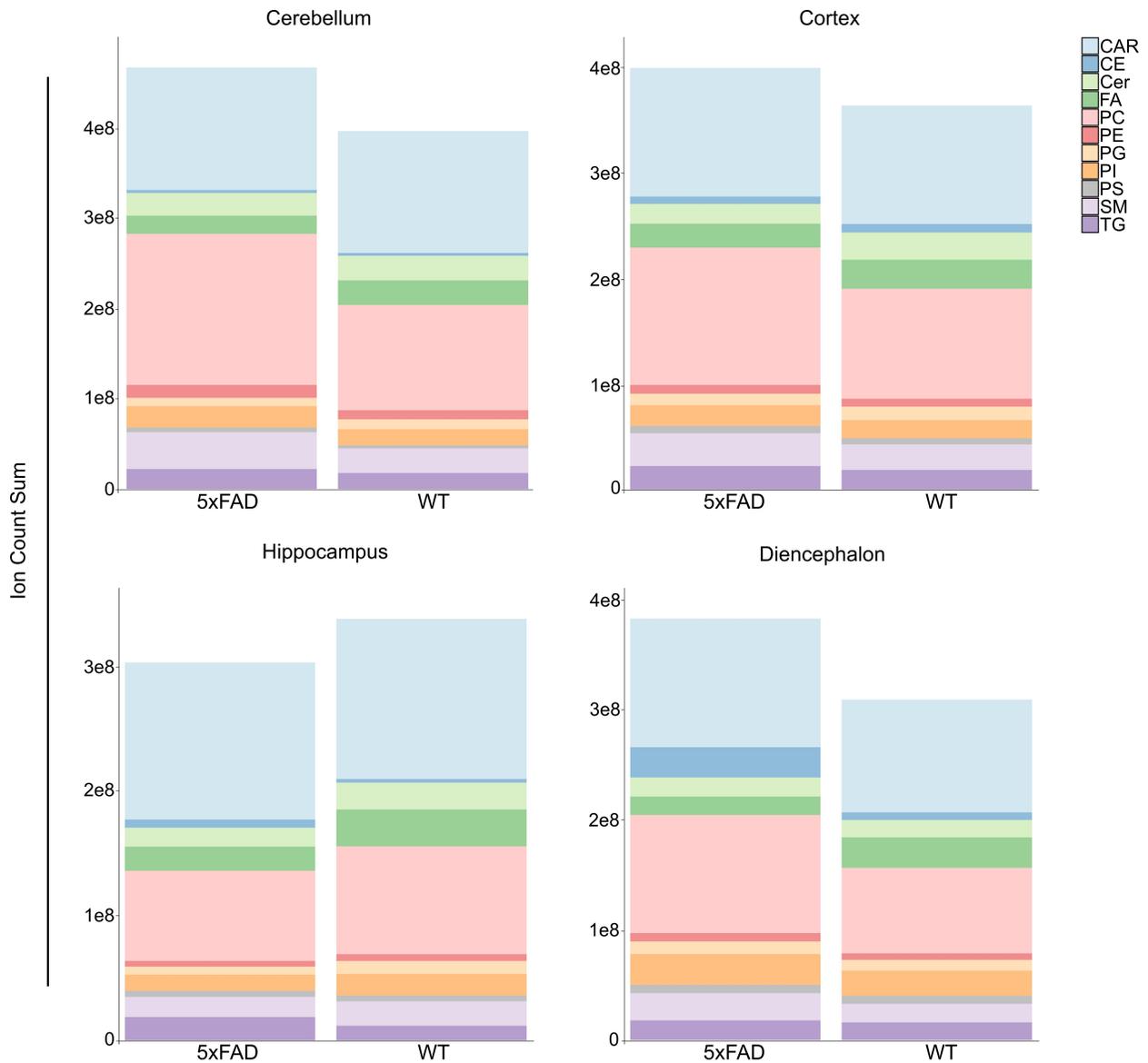

**Figure S21**. **Bar plot of summed ion counts for all lipids in 5xFAD vs WT in different brain regions: cerebellum, cortex, hippocampus, and diencephalon**. This shows that 5xFAD has higher lipid expression in lipid droplets in all regions except the hippocampus.

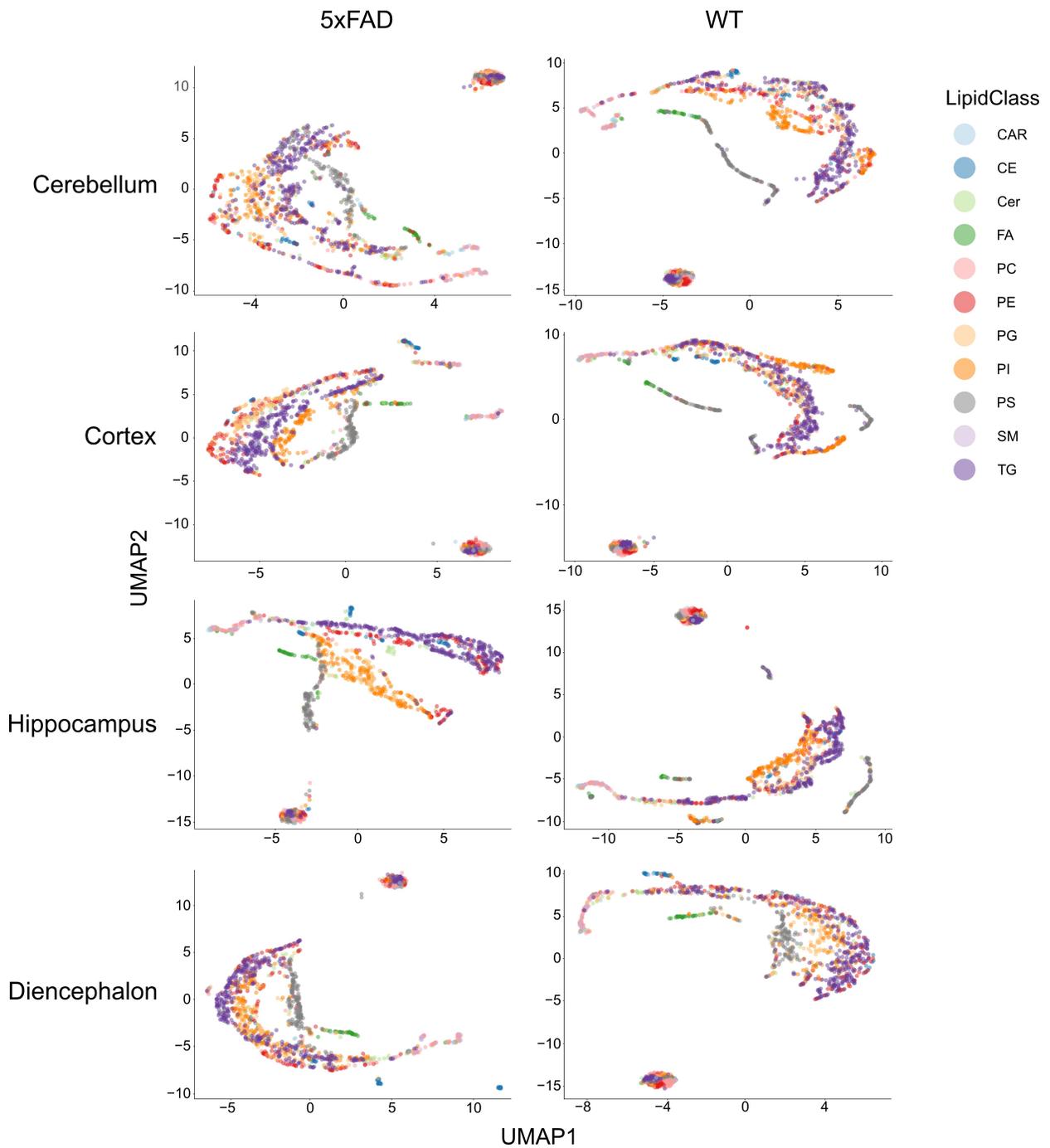

**Figure S22**. **UMAP plots of all lipids in different brain regions**: Cerebellum, cortex, hippocampus, and diencephalon. Lipid class is differentiated by color according to the legend.

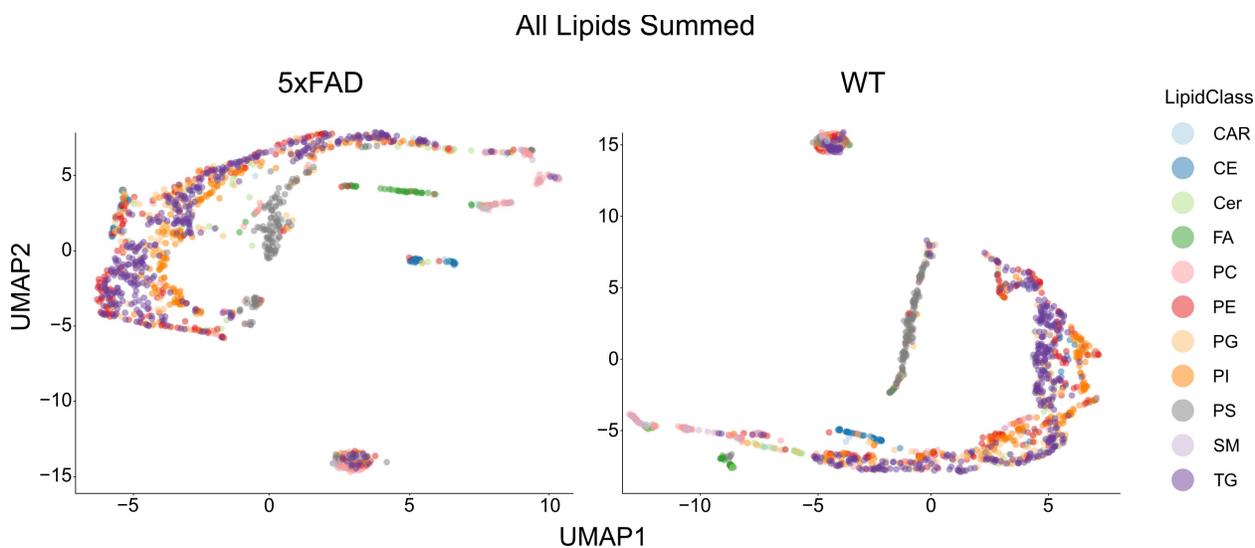

**Figure S23**. **UMAP plot of all lipids summed from four brain regions**: Cerebellum, cortex, hippocampus, and diencephalon. Lipid class is differentiated by color according to the legend.

## SUPPORTING TABLES

**Table S1** – Number of MRM transitions analyzed for each lipid class with their abbreviation and total transition count.

| Lipid Class | Abbreviation | Transition Count |
|---|---|---|
| Triacylglycerol | TG | 383 |
| Phosphatidyl Inositol | PI | 179 |
| Phosphatidyl Ethanolamine | PE | 179 |
| Phosphatidyl Serine | PS | 176 |
| Phosphatidyl Glycerol | PG | 175 |
| Phosphatidyl Choline | PC | 169 |
| Ceramide | CER | 80 |
| Cholesteryl Ester | CE | 58 |
| Fatty Acid | FA | 36 |
| Acyl Carnitine | CAR | 32 |
| Sphingomyelin | SM | 30 |

**Table S2** – Significant lipids in across brain regions cortex, cerebellum, hippocampus, and diencephalon combined of 5xFAD vs WT male mice.

| Lipid | Class | logFC | PValue | FDR |
|---|---|---|---|---|
| C24:6 | FA | -2.54 | 1.33E-21 | 2.00E-18 |
| 22:3 Campesteryl ester | CE | 1.52 | 8.13E-08 | 6.08E-05 |
| C24:5 | FA | -1.38 | 1.35E-07 | 6.72E-05 |
| 22:2 Campesteryl ester | CE | 1.45 | 2.78E-07 | 1.04E-04 |
| 22:0 Cholesteryl ester, 20:0 Sitosteryl ester | CE | 1.18 | 2.46E-05 | 0.005661 |
| 22:1 Cholesteryl ester, 20:0 Stigmasteryl ester, 20:1 Sitosteryl ester | CE | 1.19 | 1.93E-05 | 0.005661 |
| Cholesteryl 11-hydroperoxy-eicosatetraenoate, 22:2 Cholesteryl ester, 20:1 Stigmasteryl ester, 20:2 Sitosteryl ester | CE | 1.17 | 2.65E-05 | 0.005661 |
| Lyso_PE(18:4) | PE | 1.13 | 5.01E-05 | 0.009371 |
| PG (32:5) | PG | -0.97 | 2.12E-04 | 0.035298 |
| ecdysone palmitate, 22:3 Cholesteryl ester, 20:2 Stigmasteryl ester, 20:3 Sitosteryl ester | CE | 0.95 | 5.50E-04 | 0.082372 |
| PG (32:5) | PG | -3.43 | 9.62E-09 | 0.00 |
| C24:6 | FA | -2.23 | 1.26E-04 | 0.06 |
| PS (32:5) | PS | -2.27 | 9.48E-05 | 0.06 |

**Table S3** – Significant lipids in brain region cerebellum of 5xFAD vs WT male mice

| Lipid | logFC | PValue | FDR | Class |
|---|---|---|---|---|
| C24:6 | -2.64 | 7.74E-06 | 0.01 | FA |

**Table S4** – Significant lipids in brain region cortex of 5xFAD vs WT male mice

| Lipid | logFC | PValue | FDR | Class |
|---|---|---|---|---|
| C24:6 | -2.47 | 1.65E-05 | 0.02 | FA |

**Table S5** – Significant lipids in brain region diencephalon of 5xFAD vs WT male mice

| Lipid | Class | logFC | PValue | FDR |
|---|---|---|---|---|
| C24:6 | FA | -2.96 | 4.86E-09 | 7.28E-06 |
| 22:3 Campesteryl ester | CE | 2.41 | 1.94E-05 | 0.01 |
| 22:2 Campesteryl ester | CE | 2.34 | 3.23E-05 | 0.02 |
| 22:0 Cholesteryl ester, 20:0 Sitosteryl ester | CE | 2.24 | 6.00E-05 | 0.02 |
| 22:1 Cholesteryl ester, 20:0 Stigmasteryl ester, 20:1 Sitosteryl ester | CE | 2.26 | 5.28E-05 | 0.02 |
| Cholesteryl 11-hydroperoxy-eicosatetraenoate, 22:2 Cholesteryl ester, 20:1 Stigmasteryl ester, 20:2 Sitosteryl ester | CE | 2.20 | 8.08E-05 | 0.02 |
| PS (32:5) | PS | 2.17 | 9.88E-05 | 0.02 |
| ecdysone palmitate, 22:3 Cholesteryl ester, 20:2 Stigmasteryl ester, 20:3 Sitosteryl ester | CE | 1.97 | 3.53E-04 | 0.07 |

**Table S6** – Significant lipids in brain region Hippocampus of 5xFAD vs WT male mice

| Lipid | logFC | PValue | FDR | Class |
|---|---|---|---|---|
| PG (32:5) | -3.43 | 9.62E-09 | 1.44E-05 | PG |
| C24:6 | -2.23 | 1.26E-04 | 0.06 | FA |
| PS (32:5) | -2.27 | 9.48E-05 | 0.06 | PS |

**Table S7** – CLAW OzESI max intensity ratios for canola oil at the crude, degummed, and RBD filtration steps.

| Lipid | RT (min) | Crude | Degummed | RBD |
|---|---|---|---|---|
| [TG(52:2)]_FA18:1 | 18.05 | 4.23 | 4.31 | 4.70 |
| [TG(52:3)]_FA18:1 | 16.09 | 3.16 | 2.50 | 3.07 |
| [TG(52:4)]_FA18:1 | 14.30 | 3.05 | 2.96 | 2.68 |
| [TG(54:2)]_FA18:1 | 20.00 | 4.31 | 3.84 | 4.48 |
| [TG(54:3)]_FA18:1 | 17.98 | 5.81 | 5.75 | 5.41 |
| [TG(54:4)]_FA18:1 | 16.07 | 4.91 | 4.07 | 4.97 |
| [TG(54:5)]_FA18:1 | 14.31 | 4.09 | 3.83 | 3.83 |

**Table S8** – Manually processed OzESI area ratios for canola oil in the crude, degummed, and RBD filtration steps

| Lipid | RT (min) | Crude | Degummed | RBD |
|---|---|---|---|---|
| [TG(52:2)]_FA18:1 | 18.05 | 3.92 | 3.76 | 4.13 |
| [TG(52:3)]_FA18:1 | 16.09 | 2.20 | 3.25 | 2.34 |
| [TG(52:4)]_FA18:1 | 14.30 | 2.31 | 2.25 | 2.18 |
| [TG(54:2)]_FA18:1 | 20.00 | 4.27 | 3.03 | 4.36 |
| [TG(54:3)]_FA18:1 | 17.98 | 4.88 | 4.94 | 4.57 |
| [TG(54:4)]_FA18:1 | 16.07 | 3.78 | 4.06 | 4.04 |
| [TG(54:5)]_FA18:1 | 14.31 | 5.58 | 5.65 | 4.45 |

**Table S9** – Standard deviation across purities crude, degummed, and RBD for CLAW max intensity ratios and manual area ratios

| Lipid | CLAW SD | Manual SD |
|---|---|---|
| [TG(52:2)]_FA18:1 | 0.25 | 0.19 |
| [TG(52:3)]_FA18:1 | 0.36 | 0.57 |
| [TG(52:4)]_FA18:1 | 0.19 | 0.07 |
| [TG(54:2)]_FA18:1 | 0.33 | 0.74 |
| [TG(54:3)]_FA18:1 | 0.22 | 0.20 |
| [TG(54:4)]_FA18:1 | 0.50 | 0.16 |
| [TG(54:5)]_FA18:1 | 0.15 | 0.67 |
| Average | 0.29 | 0.37 |

**Table S10** – Standard deviation for the difference between CLAW max intensity vs manual area ratios between lipids for each purity

| Lipid | SD |
|---|---|
| Crude | 0.84 |
| Degummed | 0.92 |
| RBD | 0.50 |
| Average | 0.75 |

## SUPPORTING MEDIA & DATA FILES

**Video S1** – User interaction with LUI to create two JSON files for comparison of two brain regions, cortex vs cerebellum.
**Video S2** – Showing robustness of LUI. User misspells multiple words and asks poorly structured questions. LUI is able to understand and create the correct JSON files requested by the user for analysis.
**Data S1** – Brain region MRM results. Excel file with the lipid, class, MRM transition, intensity, sample ID, genotype, brain region, and sex.
**Data S2** – Canola Oil OzESI results. Excel file with the lipid, class, MRM transition, OzESI max intensity, sample ID, double bond position, and ratio of n-9/n-7 max intensity.